\documentclass[iop,apj,numberedappendix,nofootinbib]{emulateapj}
\usepackage{graphicx,amsmath,amssymb,color,bm,natbib}
\usepackage[]{hyperref}
\def\be{\begin{equation}}
\def\ee{\end{equation}}
\def\bea{\begin{eqnarray}}
\def\eea{\end{eqnarray}}
\newcommand{\vs}{\nonumber\\}
\def\ba#1\ea{\begin{align}#1\end{align}}
\def\bg#1\eg{\begin{gather}#1\end{gather}}

\newcommand{\veta}{\bm{\eta}}

\newcommand{\g}{\gamma}
\newcommand{\s}{\sigma}
\newcommand{\refeq}[1]{Eq.~(\ref{eq:#1})}          
\newcommand{\refeqs}[2]{Eqs.~(\ref{eq:#1})--(\ref{eq:#2})}          
          
\newcommand{\reffig}[1]{Fig.~\ref{fig:#1}}          
          
\newcommand{\refsec}[1]{Sec.~\ref{sec:#1}}          
\newcommand{\refapp}[1]{App.~\ref{app:#1}}

\def\N{\mathcal{N}}

\renewcommand{\v}[1]{\mathbf{#1}}

\newcommand{\vx}{\v{x}}

\newcommand{\<}{\langle}
\renewcommand{\>}{\rangle}

\renewcommand{\k}{\kappa}
\renewcommand{\d}{\delta}
\newcommand{\D}{\Delta}

\newcommand{\vnhat}{\v{\hat{n}}}

\newcommand{\eps}{\epsilon}

\def\rPSF{r_{\rm PSF}}

\newcommand{\footnoteref}[1]{\footnote{\href{#1}{#1}}}

\def\ngal{several hundred~}

\def\vnhat{\hat{\v{n}}}

\begin{document}

\title{Dust Content, Galaxy Orientations, and Shape Noise in Imaging Surveys}

\author{Petchara Pattarakijwanich}
\affiliation{Department of Astrophysical Sciences, Princeton University,
Princeton, NJ~08544, USA}

\author{Fabian Schmidt}
\affiliation{Max-Planck-Institute for Astrophysics, D-85748 Garching, Germany}

\begin{abstract}
We show that dust absorption in disk galaxies leads to a color- and
orientation-dependent centroid shift which is expected to be observable
in multi-band imaging surveys.  This centroid shift is an interesting
new probe which contains astrophysically and cosmologically relevant
information: it can be used to probe the 
dust content of a large sample of galaxies, and to reduce the shape noise
due to inclination of disk galaxies for weak lensing shear.  
Specifically, we find that data sets comparable to CFHTLenS, the
Dark Energy Survey (DES) or the Hyper Suprime-Cam (HSC) survey should provide a dust measurement for
\ngal galaxies per square degree.  Conversely, given knowledge of the
dust optical depth, this technique will significantly lower the shape
noise for the brightest galaxies in the sample (signal-to-noise greater
than a few hundred), thereby increasing their relative importance for the
weak lensing shear measurement.
\end{abstract}

\maketitle

\section{Introduction}
\label{sec:intro}

Observational cosmology is currently benefiting from an enormous growth
in large photometric survey data sets, as delivered by the CFHTLenS\footnoteref{http://www.cfhtlens.org/}, 
Hyper Suprime-Cam (HSC)\footnoteref{http://www.subarutelescope.org/Projects/HSC/} survey, Dark Energy Survey (DES)\footnoteref{http://www.darkenergysurvey.org/}, KiDS\footnoteref{http://kids.strw.leidenuniv.nl/} and other surveys.  
These state of the art surveys deliver images in several bands for a hundred
million galaxies or more.
Currently, these data sets are exploited primarily through three observables:  
image shapes, which are used to measure weak gravitational lensing shear; 
galaxy number counts, used to measure angular auto- and cross-correlation
functions; and image magnitudes in different bands, which are the input
for photometric redshift estimation.  

In this paper, we show that the image centroid measured in different
bands can also serve as a valuable source of cosmological information.  
In particular, it can be used to infer the dust content of galaxies,
and to infer the orientation of disk galaxies independently of their
observed shape.  The underlying idea is that dust absorption modifies
galaxy images by a different amount in different bands.  The most easily
observable of these modifications is a \emph{shift in the centroid} of the image
between different bands.  As we will show, a radial gradient in the
disk plane of the optical depth is sufficient to produce such a centroid
shift, which is approximately proportional to the tangent of the inclination
angle.

Direct measurements of the optical depth due to dust absorption 
through the disk of spiral galaxies has mainly
been done using two techniques. The first is to observe occulting galaxy pairs,
where comparing the unobscured part to the obscured part of the background
galaxy can give information on the transparency of the foreground one
(e.g., \citealt{white/etal:00}). The second approach is through the number count of
distant background galaxies observed through the disk of local face-on galaxies
(e.g., \citealt{holwerda/etal:13}).  Both of these methods rely on spatially resolved
observation, limiting the number of galaxies studied this way to only tens in
the local universe.

There are other ways to indirectly infer the dust optical depth in galaxies.
Far-IR spectral energy distributions from Spitzer and Herschel can be used to
measure total dust emission, implying the total dust mass and expected
obscuration in the optical. However, relating the FIR SED to optical depth in
visible wavelength is not straightforward due to the uncertainty in dust models
(e.g., \citealt{draine/etal:07}). Dust content in late-type galaxies can be determined by
detailed modeling of photometric and spectroscopic data in small samples of
nearby galaxies (see \citealt{calzettireview} for a review). A similar approach that
works in a statistical sense (i.e. averaged over many galaxies) with spatially-unresolved data is to fit
stellar populations using optical spectra, and then infer dust reddening from
the discrepancy between the observed photometry and the best-fit model
(e.g., \citealt{kauffmann/etal:03}). Even though this can be done for a large number of
galaxies from surveys, the uncertainty is large due to the systematics in stellar
population modeling.

Effects of inclination-dependent obscuration have been considered by a number
of authors. \citet{shao/etal:07} showed that the luminosity function of spiral
galaxies depends on inclination in the expected way, then determined the
inclination-corrected luminosity function. \citet{chevallard/etal} determined
the dust content of nearby star-forming galaxies using fits to the optical/NIR
photometry with sophisticated radiative transfer model taking this effect
into account. \citet{yip/etal:11} considered the possible biases induced by
inclination-dependent reddening on derived photometric redshift.

The centroid shift we describe in this paper is an important new observable
which is able to break degeneracies present when simultaneously estimating the
dust content and galaxy orientation.  Thus, by using this technique in
conjunction with the previously developed ones, one can expect to obtain
estimates for the dust that are significantly more robust.  

The degeneracy between an elliptical appearance due
to inclination of a disk galaxy and the ellipticity induced by weak
lensing along the line of sight is the main source of noise in shear
measurements.  Thus, any independent constraint on the orientation of
a disk galaxy can increase the signal-to-noise of shear measurements.  
This is especially relevant since disk galaxies typically make up the
majority of galaxies in flux-selected samples from imaging surveys.  
We show how (noisy) information on the orientation, in our case from the
centroid shift, can be included in
the shear estimate from galaxy shapes.  This could also be useful in
other contexts, for example when incorporating orientation
information from spectra \citep{huff/etal:14}.  

The outline of the paper is as follows.  In \refsec{centroid}, we 
present the calculation of the centroid shift due to dust absorption in
disk galaxies.  We perform both a numerical solution of the radiative
transfer problem as well as an analytical estimate.  \refsec{improveshear}
then shows how the information on the galaxy orientation can be incorporated
in shear estimation to lower the shape noise.  We then briefly describe the
constraints on the dust content that can be obtained via this technique.  
We conclude in \refsec{concl}.  The appendices present further details
on analytical estimates and the expected accuracy of centroid measurements
in imaging surveys.

\section{Centroid shift from dust extinction}
\label{sec:centroid}

In this section, we use a simplified radiative transfer calculation to show
why dust extinction in disk galaxies leads to a relative
centroid shift between different bands.  For this, we will ignore scattering. 
The radiative transfer equation then simplifies to
\be
\frac{dI_\nu(\vx(s),\vnhat)}{ds} = - \k_\lambda(\vx(s)) I_\nu(\vx(s),\vnhat) + \eps_\lambda(\vx(s), \vnhat)\,,
\label{eq:radtrans}
\ee
where $I_\nu$ is the photon intensity, $\vnhat$ is the unit photon momentum
vector, $s$ is the line element, $\vx(s)$ is the position along the ray, and
$\k_\lambda,\,\eps_\lambda$ are the opacity and emissivity, respectively. In
the following, we will drop the subscript $\lambda$ for clarity. We will assume an
isotropic opacity $\k$ and emissivity $\eps$, and correspondingly drop the argument $\vnhat$
throughout. The solution to \refeq{radtrans} is then given by
\be
I(s) = \int_{-\infty}^s ds' \: \eps(\vx(s')) \exp\left[-\int_{s'}^{s} \k(\vx(s'')) ds'' \right]\,.
\label{eq:Isol}
\ee

\begin{figure}[t!]
\centering
\includegraphics[clip=true, trim=4cm 5cm 4cm 5cm, width=0.48\textwidth]{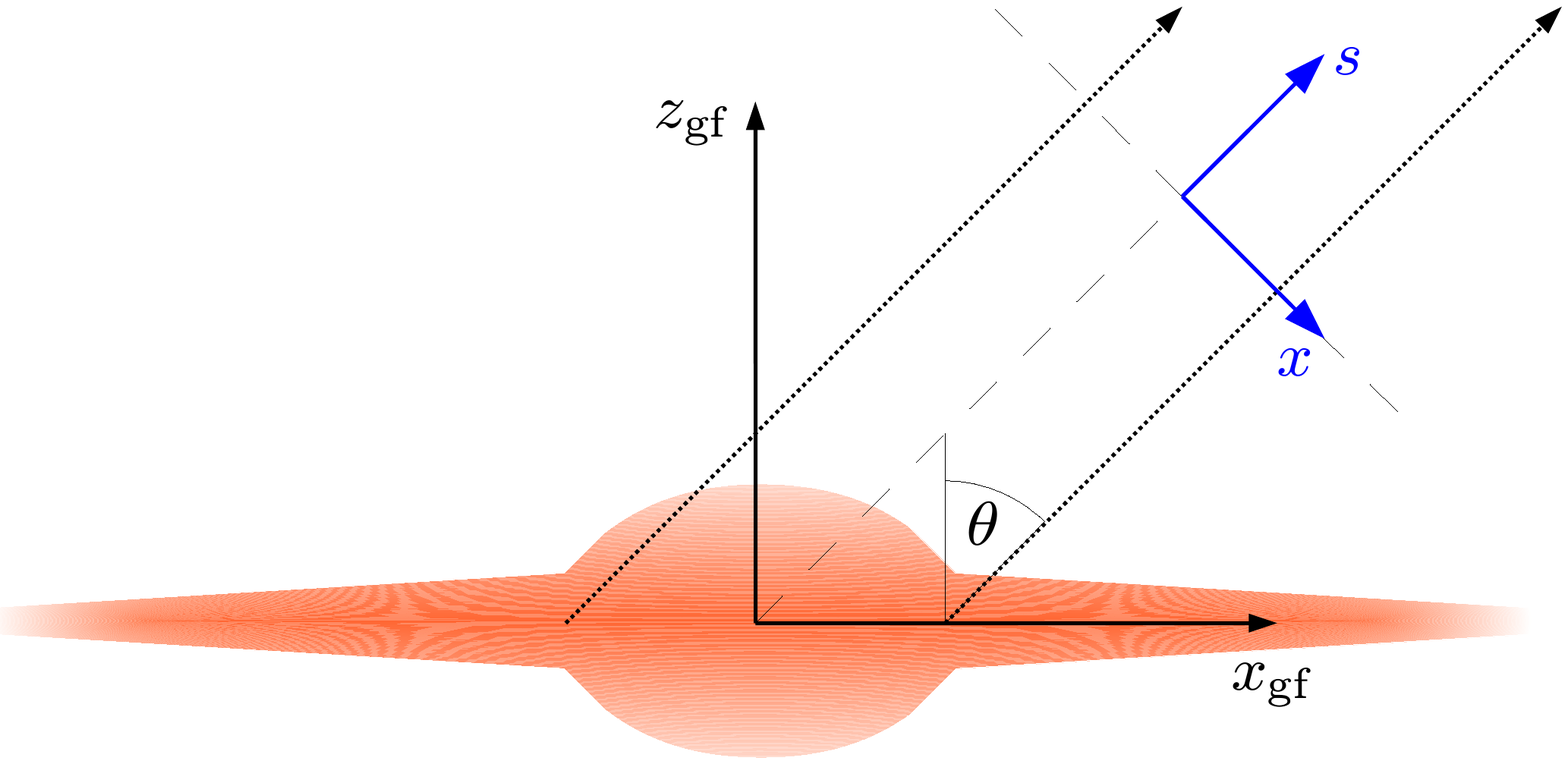}
\caption{Geometry of the galaxy and lines of sight.  The black coordinate
system shows the galaxy frame (``gf'') while the blue coordinate axes
denote the observer system $(x,y)$ on the sky plane (the $y$ axes of both
coordinate systems are perpendicular to the figure).  $\theta$ is the inclination
angle, and we have set $\varphi=0$. The two lines of sight (dotted) at $\pm |x|$ in the observer
frame pass through different amounts of material and thus experience 
difference extinction.
\label{fig:sketch}}
\end{figure}

Let us consider a galaxy with cylindrical symmetry, so that in cylindrical
coordinates around the galaxy center [``galaxy frame'', $(r_{\rm gf},z_{\rm gf},\varphi)$] we can write $\eps(\vx) = \eps(r_{\rm gf},z_{\rm gf})$
and $\k(\vx) = \k(r_{\rm gf},z_{\rm gf})$. Consider a ray that intersects the $z_{\rm gf}=0$ plane
(``the disk'', see \reffig{sketch}) at $(x_{\rm gf}, y_{\rm gf})$, which we define as
$s=0$, at an angle $\theta$ with the $z_{\rm gf}$ axis.  The angle $\theta$ corresponds
to the inclination angle as seen by a distant observer, with $\theta=0$ corresponding to a face-on view and $\theta=\pi/2$ being edge-on.  This yields
\begin{equation}
z_{\rm gf}(s) = s \cos\theta;\quad
r_{\rm gf}(s) = \left((x_{\rm gf} + s \sin\theta)^2 + y_{\rm gf}^2\right)^{1/2}\,.
\end{equation}
We can replace $s$ with $z$ (provided that $\theta \neq \pi/2$). Further,
we are interested in the flux $f_{\rm obs} \propto I_{\rm obs}$ far away from 
the galaxy, and thus let $s\to\infty$. \refeq{Isol} becomes
\begin{align}
 I_{\rm obs}(\theta,x_{\rm gf},y_{\rm gf}) &= \frac1{\cos\theta} \int_{-\infty}^\infty dz_{\rm gf} \: \eps(r_{\rm gf}(z_{\rm gf}), z_{\rm gf}) \label{eq:Isolobs}\\
&\times \exp\left[-\frac1{\cos\theta}\int_{z}^{\infty} \k(r_{\rm gf}(z'), z') dz' \right]\,,
\nonumber
\end{align}
where
\be
r_{\rm gf}(z_{\rm gf}) = [(x_{\rm gf}+z_{\rm gf} \tan\theta)^2+y_{\rm gf}^2]^{1/2}\,.
\ee
Evaluating this expression yields the appearance of the disk with
inclination angle $\theta$ as seen by a distant observer.  For this purpose,
we define a coordinate system on the sky $(x,y)$, assuming the flat-sky approximation, centered on the true
center of the galaxy and scaled with the angular diameter distance to the
galaxy.  The mapping between ($x,y$) and $(x_{\rm gf}, y_{\rm gf})$ then is
\be
\left(\begin{array}{c}
x_{\rm gf} \\
y_{\rm gf}
\end{array}\right)
= \left(\begin{array}{c}
x \cos\theta\cos\varphi_i - y \cos\theta \sin\varphi_i \\
y \cos\varphi_i + x \sin\varphi_i
\end{array}\right)\,.
\ee
where $\varphi_i$ is the azimuthal angle of the inclination axis.

\begin{figure}[t!]
\centering
\includegraphics[width=0.45\textwidth]{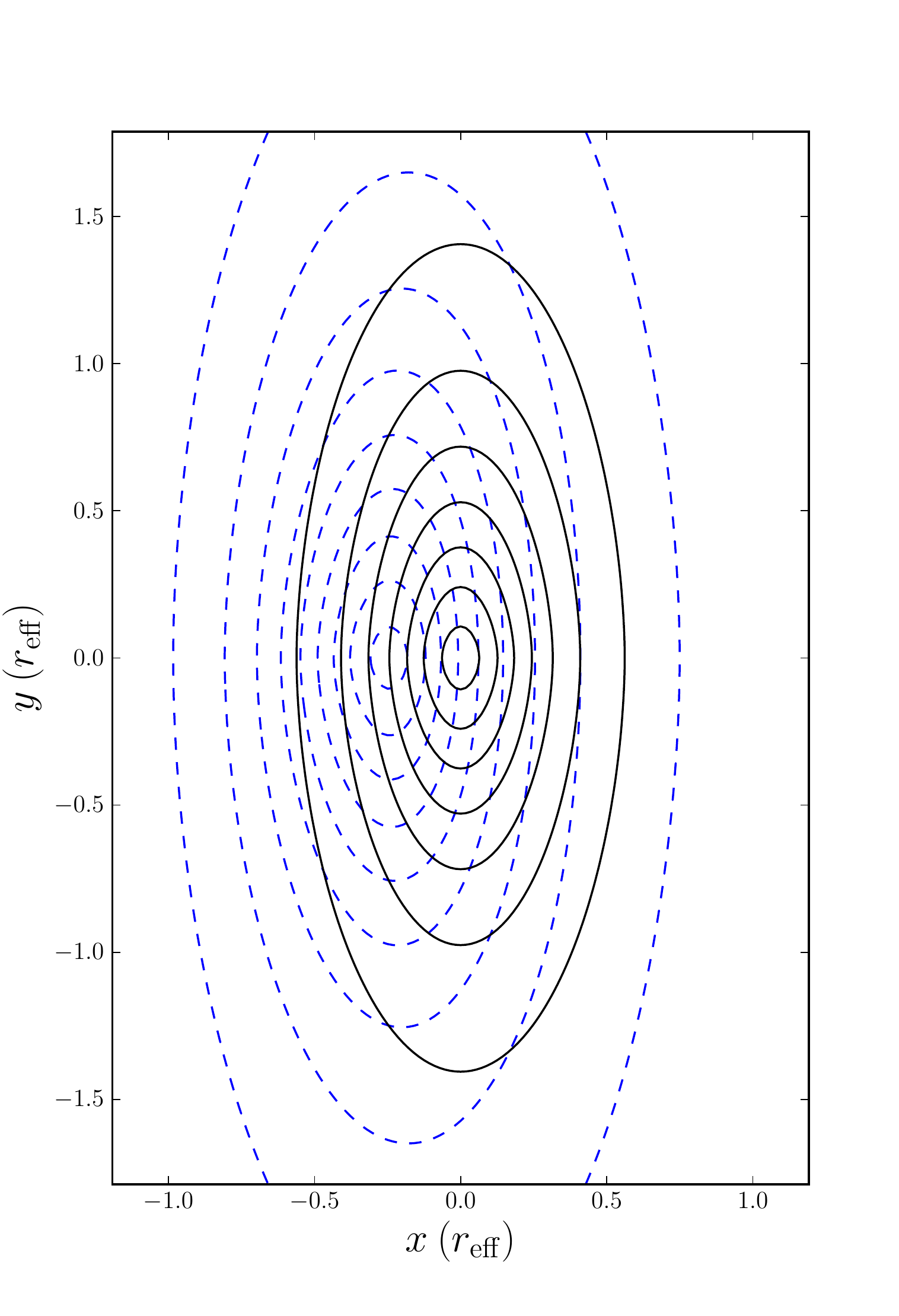}
\caption{Contours showing the appearance of the disk with (blue dashed) and without (black solid) the
effect of dust obscuration.  Inclination angle
here is $\theta=70^\circ$. The disk parameters used here are the fiducial set
of parameters [\refeq{diskpar}] except for the optical depth which is boosted from $\tau_{\rm fo}=0.5$ to
$5.0$ to show the effect clearly.  The axes are scaled to the effective
radius $r_{\rm eff}$ (half-light radius) for a face-on view.}
\label{fig:compare_contour}
\end{figure}

\subsection{Numerical Evaluation}

In the following, we perform a numerical integration of the radiative
transfer solution \refeq{Isolobs} for the case when
both the absorbing and emitting components of the galaxy follow a
double exponential distribution in $r_{\rm gf}$ and $z_{\rm gf}$,
\begin{align}
\eps(r_{\rm gf},z_{\rm gf}) &= \eps_0 e^{-r_{\rm gf}/r_\eps} e^{-|z_{\rm gf}|/z_\eps} \vs
\k(r_{\rm gf},z_{\rm gf}) &= \k_0 e^{-r_{\rm gf}/r_\k} e^{-|z_{\rm gf}|/z_\k}\,.
\label{eq:exponentialdisk}
\end{align}
Assuming this model, \refeq{Isolobs} can be evaluated numerically for any set
of parameters.  The emissivity normalization $\eps_0$ and the absolute scale
length $r_\eps$ do not need to be specified, since they only scale the solution
in a trivial way. The relevant quantities are the opacity $\k_0$ and the ratios
of the scale lengths and heights. We estimate the value of $\k_0$ from the
face-on optical depth, $\tau_{\rm fo}$, which is the optical depth corrected
for inclination and averaged over the inner $2r_\eps$
(approximately one effective radius, consistent with how 
it is measured in the literature , see \citet{calzettireview} for a review).
In the following, we
assume the values of optical depths in B- and I-bands and ratios of scales to
be
\begin{align}
\tau_{\rm fo,B} \simeq\:& 0.5 \vs
\tau_{\rm fo,I} \simeq\:& 0.2 \vs
z_\eps / r_\eps \simeq\:& 0.2 \vs
r_\k / r_\eps \simeq\:& 1.4 \vs
z_\k / z_\eps \simeq\:& 0.5 \,.
\label{eq:diskpar}
\end{align}
These values of optical depths are typical for Sb-Scd galaxies summarized by
\citet{calzettireview} (see however \cite{Driver/etal:07} who obtain somewhat
larger values).  Note that these values are for optically-selected
low-redshift galaxies and there is evidence that infrared-selected galaxies and
galaxies at higher-redshift have significantly larger optical depths
\citep{Grootes/etal:13, Boquien/etal:13, Sargent/etal:10}. Therefore it is
well possible that the galaxies detected in deep imaging surveys will show larger
values of $\tau_{\rm fo}$. The values used for ratios of scale lengths and
heights are representative of what is found in the literature, where they are derived
from analysis of edge-on local spiral galaxies \citep{Byun/etal:94,
Xilouris/etal:99, Bianchi:07, Gadotti/etal:10}. Despite the significant
variation from galaxy to galaxy, all measurements
consistently suggest that $z_\k/z_\eps < 1$, i.e., the source of attenuation
is more concentrated to the disk than the emission.  We show in \refsec{analytical}
that this ratio governs the sign of centroid shift, and therefore our fiducial
model with fixed parameters is still qualitatively robust despite possible variations in
these parameters.

With the surface brightness at each position $(x,y)$ on the sky-plane projection of the disk calculated,
one can explore how the extinction affects the appearance of the disk in
different bands. The appearance of the disk in one band is shown in \reffig{compare_contour}. The solid black contour shows the disk if no dust is
present, while the blue dashed contour includes the effect of dust extinction.
The disk shown in this figure is at inclination angle $\theta=70^\circ$ with
our fiducial set of parameters except for the dust optical depth, which is
boosted by a factor of 10 to $\tau_{\rm fo}=5.0$ in order to show
the effect more clearly.

\begin{figure}[t!]
\centering
\includegraphics[width=0.5\textwidth]{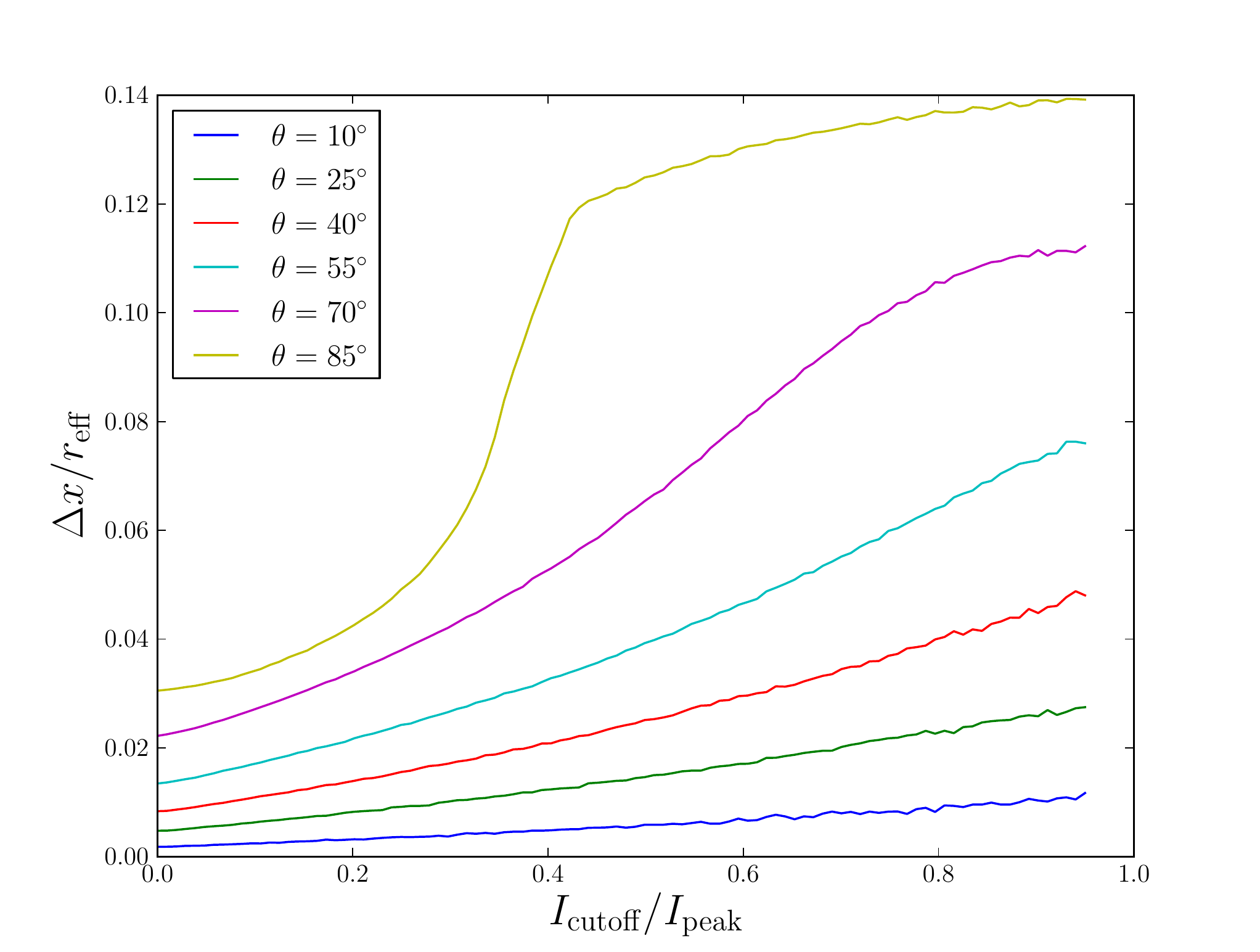}
\caption{The magnitude of the fractional centroid shift $\Delta x/r_{\rm eff}$ 
as a function of cutoff surface brightness (in units of the peak value at the galaxy center) of the region used to measure the centroid. The different lines show
the same disk from different inclination angles. The disk parameters are 
given in \refeq{diskpar} with $\tau_{\rm fo} = 0.5$.}
\label{fig:shift_threshold}
\end{figure}

The effect of extinction shifts the disk systematically to one side, as
expected because the radiation from one side goes through thicker column of
absorber than the other.  As dictated by symmetry, the shift is perpendicular
to the axis around which the galaxy is rotated.  
The amount of shifting, however, is not uniform across
the disk but depends on each surface brightness level. The bright, central part
is shifted by a larger fractional amount than the fainter part. This is a consequence of the
larger extinction gradient toward the center of the galaxy.  To explore this
effect, we divide the simulated disk image into isophotes (contours of constant
surface brightness), and plot the shift of isophotes above some surface
brightness threshold as a function of threshold value (see \reffig{shift_threshold}).  
Here and throughout, we scale the centroid shift to the effective radius 
$r_{\rm eff}$, or half-light radius, the radius within which one half of 
the total luminosity of the galaxy is emitted.  
This radius is related to the radial scale length $r_\eps$ approximately by 
$r_{\rm eff} \sim 1.7r_\eps$.  
As the threshold increases, we only calculate the centroid from the brightest 
part of the disk, and the centroid
shift grows as expected. For the almost edge-on disk, a strongly non-linear 
behavior is apparent.

\begin{figure}[t!]
\centering
\includegraphics[width=0.49\textwidth]{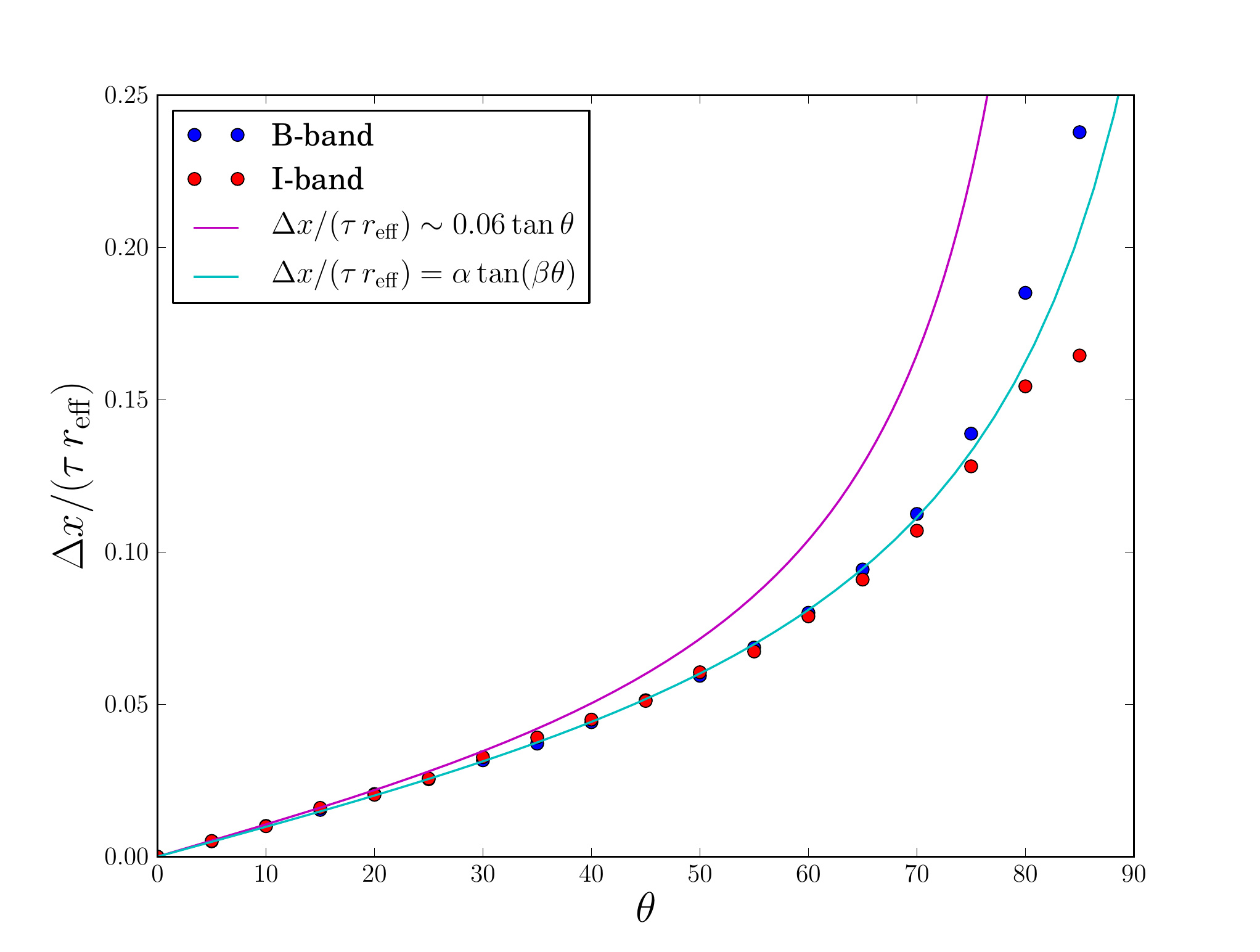}
\caption{Fractional centroid shift calculated for both the $B$-band ($\tau_{\rm fo}=0.5$,
blue point) and the $I$-band ($\tau_{\rm fo}=0.2$, red point). The shifts are normalized
by the optical depth in each respective band to show the linear dependence.  
The magenta line is the analytic approximation discussed in \refsec{analytical}, which agrees with the full
numerical evaluation at small inclinations as expected.  
The cyan line is the functional fit to both sets of points [\refeq{cent_param}], which we used as
the fiducial relation between the shift and inclination angle.}
\label{fig:shift_angle}
\end{figure}

In \reffig{shift_angle}, we show the centroid shift, measured within $0.5r_{\rm eff}$
(for reasons fully explained in \refsec{centroid_params}) scaled to the effective
radius $r_{\rm eff}$ and the optical depth $\tau_{\rm fo}$, as a function of inclination angle.  We see
that the centroid shift can be well approximated as linear in $\tau_{\rm fo}$.  
The two lines show the analytical approximation presented in the next
section, and a parametrization which we will use for the following results.

\subsection{Analytical estimates}
\label{sec:analytical}

The physical reason for the centroid shift can be inferred from \reffig{sketch},
where we have let the $y$ axis (perpendicular to the page) be the orientation
axis of the galaxy.  A ray emitted towards the observer from the negative
$x$ axis passes through matter closer to the center than the opposite ray
from the positive $x$ axis.  This means that it will experience more emission 
and absorption than the latter;  which contribution dominates depends
on the configuration of absorbing and emitting material.  Since the
$x>0$ half of the galaxy image now has a different flux than the $x<0$
half, the centroid of the observed image, i.e. the center-of-mass of the 
light distribution, is shifted from the $x=0$
position of the actual galaxy center.  
Of course, for $\theta=0$ both rays pass through
equivalent emission and absorption (on average), and there will be no 
centroid shift.  More generally, both a radial and $z$-gradient is necessary
in \emph{either} the opacity or emissivity.  

An analytical calculation of the centroid shift can be performed for small
inclination angle, by expanding \refeq{Isolobs} to linear order in 
$\tan\theta$.  The details of this are given in
Appendix \ref{app:smallangle}.  For the fiducial parameters
\refeq{diskpar}, the centroid shift is along the negative $x$-axis, i.e.
the effect of more emission at smaller radii wins over the higher
absorption.  This is due to the more concentrated distribution of
opacity as compared to the emissivity; while the former is dominated by
gas clouds largely confined to the disk, the latter also receives contributions
from older stellar populations which are more extended.  Quantitatively,
the analytical estimate yields [\refeq{xcf}]
\be
\frac{\D x_{\rm cen}}{r_{\rm eff}} \simeq -0.06
\tan\theta\: \tau_{\rm fo}\,.
\label{eq:shift_params}
\ee
This linear approximation is compared to the
direct numerically evaluated result in \reffig{shift_angle}.  
For inclination angles less than about 45 degrees, the numerical results
follow $\tan\theta$ closely.  Moreover, the coefficient of the analytical
estimate, which uses some simplifications in various weighting integrals, 
agrees with the numerical result to better than 10\%.

\subsection{Parametrization of centroid shift}
\label{sec:centroid_params}

For the purpose of forecasting how well the centroid shift can be measured in
imaging surveys, we need a parametrization of the centroid shift as a function
of optical depth $\tau_{\rm fo}$ and inclination $\theta$. Since the effect depends on
isophote level (\reffig{shift_threshold}), the first step is to devise
an optimal centroid estimator for the purpose of measuring the centroid shift,
which is described by a radial weight function $w(r)$. This derivation is given
in Appendix \ref{app:centroid_estimate}. It turns out that this kernel is
strongly peaked at small radii, so that, for all but the most highly resolved
galaxies, most of the information on the centroid shift is contained in the
innermost resolved region around the centroid of the galaxy, i.e. within
one full-width-half-maximum radius $\rPSF$ of the point-spread function (PSF).  

We then measure the centroid shift within $\rPSF$ in the simulated galaxy image
as a function of inclination. In a realistic galaxy survey, one has to take
into account that differently resolved galaxies will have different centroid
shifts. For the sake of simplicity, and to conservatively estimate the effect
for galaxy samples used for weak lensing, we will consider a poorly-resolved
galaxy with $r_{\rm eff} = 2 \rPSF$ in the following. This is a conservative
assumption since a better-resolved galaxy will yield a larger centroid shift.

In order to provide an accurate description even for very inclined galaxies, we
use a fit function to describe the dependence on $\theta$. We found that the
magnitude of the fractional centroid shift is well fitted by a function of the
form
\begin{align}
\frac{|\D x_{\rm cen}|}{r_{\rm eff}} =\:& F(\tau_{\rm fo}, \theta) \vs
F(\tau_{\rm fo},\theta) =\:& \alpha\,\tau_{\rm fo}\,\tan(\beta\theta)\,,
\label{eq:cent_param}
\end{align}
where $\alpha=0.066$ and $\beta=0.85$, which is a slight generalization of
the analytical result given in the previous section.  
The fit function is compared with the numerical result in 
\reffig{shift_angle} 
and is found to represent the centroid shifts well in both bands.

\subsection{Measuring orientation}

With the relationship between inclination angle and the centroid shift
calibrated, it is now possible to use this information to measure the
orientation of a disk from images in two bands (called band $a$ and $b$
respectively, which in this case our fiducial model parameters are from
$B$-band and $I$-band).  This orientation measurement assumes that
the optical depth is known for the given galaxy; the converse case,
i.e., inferring the optical depth from the centroid shift when assuming a
certain orientation, is considered in \refsec{dust}. 
Further, we also assume that there is no scatter in the optical
depths and
scale ratios [\refeq{diskpar}] that control the centroid shift at fixed
inclination. In reality, there can be significant variation in these
parameters and the empirical relation calculated in section
\ref{sec:centroid_params} can be thought of as a median relationship.

Consider a disk galaxy with an inclination angle $\theta$, and azimuthal
angle $\phi$ with respect to a fixed direction $\v{i}$ on the sky.  
We will denote positions on the sky (approximated as flat) with $\vx$ here.  
The centroid 
position in each band will be shifted from the unknown true position by 
\ba
\vx_a =\:& x_a \,\v{i} + y_a \,\v{j}
    = F(\tau_a, \theta)\, (-\cos\phi \;\v{i} + \sin\phi \;\v{j}) r_{{\rm eff}, a}\vs
\vx_b =\:& x_b \,\v{i} + y_b \,\v{j}
    = F(\tau_b, \theta)\, (-\cos\phi \;\v{i} + \sin\phi \;\v{j}) r_{{\rm eff}, b}\,,\nonumber
\ea
where $F(\tau_{\rm fo},\theta)$ is defined in \refeq{cent_param}, and 
$\tau_{a,b}$ denote the optical depths $\tau_{\rm fo}$ of the galaxy in each band.  

What we measure is the differential shift between the two bands given by
\be
\Delta\v{x} = \Delta x \;\v{i} + \Delta y \;\v{j}
    = F(\tau_a-\tau_b, \theta)\, (-\cos\phi \;\v{i} + \sin\phi \;\v{j}) r_{\rm eff}\,.
\label{eq:centr_shift}
\ee
where we have used that the shift is linear in $\tau$, and assumed that
the effective radii of the galaxy in each band are approximately equal;
the latter assumption is merely for the sake of simplicity and can be
easily dropped.  

By assuming a value for $\tau_a-\tau_b$, the angles $\theta$ and $\phi$ can then be solved for through
\begin{align}
\theta &= [F^{-1}]\left(\tau_a-\tau_b, \frac{\sqrt{\Delta x^2 + \Delta y^2}}{r_{\rm eff}} \right) \vs
       &= \frac{1}{\beta}\arctan\left(\frac{\sqrt{\Delta x^2 + \Delta y^2}}{\alpha (\tau_a - \tau_b) r_{\rm eff}}\right) \vs
\phi &= \arctan\left(-\frac{\Delta y}{\Delta x}\right)\,,
\label{eq:est_orientation}
\end{align}
where $[F^{-1}]$ denotes the inverse function (with respect to $\theta$) of
\refeq{cent_param}, given explicitly in the second line.

In order to obtain quantitative forecasts for the information contained
in the centroid shifts, we also need an estimate of the observational
uncertainty of the centroid shift measured in actual imaging surveys.  For
this, we performed a study using simulated HST imaging \citep{massey/etal:07} which is
described in detail in \refapp{sigma_centroid}.  The fractional
uncertainty in centroid position is approximately given by
\be
{\rm RMS}\left(\frac{x,\;y}{r_{\rm eff}}\right) = \frac{3.3}{\nu}\,,
\label{eq:centr_noise}
\ee
where $\nu$ is the signal-to-noise of the galaxy image.  
In other words, the uncertainty in each of the centroid
components for a galaxy measured at $\nu=10$ is roughly one quarter of the
effective radius.  We will use this estimate for the numerical studies
presented in the next section.

\section{Improving shear estimation}
\label{sec:improveshear}

We now derive how the orientation information contained in the centroid shift
can be used to reduce shape noise in weak lensing shear estimation 
of disk galaxies. This 
corresponds to an increase in the information of weak lensing shear surveys
at no observational cost, and could thus be highly relevant to ongoing
and upcoming imaging surveys.

Our goal is to construct the posterior
\be
P(\bm{\gamma} | \hat{\D\vx},\,\hat I_{\rm sky})\,,
\label{eq:post0}
\ee
where $\bm{\g} = (\g_1, \g_2)$ is the shear vector and $\hat{\D\vx},\,\hat I_{\rm sky}$
denote the measurements of the centroid shift and second moment tensor
of the galaxy image, respectively.  For the sake of simplicity, we 
will assume no noise in the
measurement of $\hat I_{\rm sky}$, since this is not the main focus of
the paper and can be straightforwardly included in the approach described
here.  In full generality, we have to allow for magnification 
as well.  We assume however that we have no knowledge about the intrinsic
size of the galaxy, so that marginalizing over the intrinsic size will remove
all information on the magnification.  To simplify the derivation, we will
thus ignore the magnification and only work with the two components of 
the shear.  Specifically, we use the shear parametrization of \cite{bernstein/jarvis},
in which the symmetric unit-determinant shear matrix is given by
\ba
S_{\veta} = R_{\varphi} \left(\begin{array}{cc}
e^{\eta/2} & 0 \\
0 & e^{-\eta/2}
\end{array}\right) R^T_{\varphi}\,,
\ea
where $\veta$ is the conformal shear vector, $\eta = |\veta|$, and $\varphi$ 
is the azimuthal angle of $\veta$. $R_{\varphi}$ denotes a rotation by $\varphi$ around the line of sight.  Note that in the small-$\eta$ limit, $\veta = 2 \bm{\gamma}$.

Consider the second moment tensor of an infinitely thin disk inclined by
an angle $\theta$, with azimuthal angle $\phi=0$.  
Normalizing this tensor to unit determinant, we obtain
\be
\frac{I_{ij}}{|I|} = \left(\begin{array}{cc}
1/\cos\theta & 0 \\
0 & \cos\theta
\end{array}\right) = S_{\veta} \v{1}_2\: S_{\veta}^T\,,
\ee
where $\v{1}_2$ is the two-dimensional identity matrix, and we have
defined the effective shear $\veta$ through
$\exp(-\eta) = \cos\theta$ and $\varphi=0$.  The
relation between $\eta$ and $\mu=\cos\theta$ for a thin disk can also be derived
by noting that the ellipticity [\refeq{edef}] is given by $e = (1-\mu^2)/(1+\mu^2)$ and
$e = \tanh \eta$ \citep{bernstein/jarvis}.  
Thus, the image of a disk galaxy oriented by angles $\theta,\phi$ is given by
the action on a face-on disk (circular image) by a shear
\be
\veta_{\rm orien}(\theta,\phi): \qquad
\exp\left[ -\eta\right] = \cos\theta; \quad \varphi = \phi\,.
\label{eq:shearorient}
\ee
Here we have chosen by convention to make $\eta$ non-negative; face-on ($\theta=0)$ corresponds to $\eta=0$, while edge-on ($\theta = \pi/2$) corresponds to $\eta\to+\infty$.  
We can thus perform the complete calculation of rotation/projection and 
lensing in conformal shear space.  For this, we need the composition law
of conformal shear, which is \emph{non-commutative}.  The
sum of two shears is given by the unique symmetric unit-determinant matrix
which is equal to the composition of the two shear matrices modulo a rotation
\citep{bernstein/jarvis}.  This yields
\ba
& \veta = \veta_1 \oplus \veta_2: \vs
&\cosh \eta = \cosh \eta_1 \cosh \eta_2 + \sinh \eta_1 \sinh \eta_2 \cos (2[\varphi_1 - \varphi_2]) \vs
& \sinh \eta \sin (2[\varphi - \varphi_1]) = \sinh \eta_2 \sin (2[\varphi_2-\varphi_1])\,.
\ea

We now turn to \refeq{post0}, written in terms of the conformal shear for the
time being.  We replace the moment tensor on the sky $\hat I_{\rm sky}$ with
the shear $\hat{\veta}$ estimated using standard techniques from the second
moments.  Using Bayes' theorem, the posterior is given by
\ba
P(\veta | \hat{\D\vx},\,\hat{\veta}) =\:& \frac1\N 
P(\hat{\D\vx},\,\hat{\veta} | \veta) P(\eta) \label{eq:post1}\\
 =\:& \frac1\N P(\eta) \int_0^1\!\! d\cos\theta \int_0^{2\pi}\!\!\!\! d\phi \: P(\hat{\D\vx},\,\hat{\veta} | \veta, \theta, \phi)\,,
\nonumber
\ea
where $\N$ is a normalization constant and $P(\eta)$ is the prior probability
of the shear $\veta$ (we assume that there is a flat prior on the angle
of the shear).  $P(\hat{\D\vx},\,\hat{\veta} | \veta, \theta, \phi)$ is the probability of obtaining a centroid shift and shear 
 as measured ($\hat{\D\vx}$ and $\hat{\veta}$, respectively),
\emph{given true}  shear vector $\veta$ and orientation
angles $\theta,\,\phi$.  We now transform the integral
over orientation angles $\theta$ to shear $\eta_d$, through \refeq{shearorient}:
\ba
& P(\veta | \hat{\D\vx},\,\hat{\veta}) 
= \frac{P(\eta)}{\N'}  \int_0^\infty e^{-\eta_{d}} d \eta_d \int_0^{2\pi} d\phi_{d} \: P(\hat{\D\vx},\,\hat{\veta} | \veta, \veta_d)
\vs
& = \frac{P(\eta)}{\N'}  \int_0^\infty e^{-\eta_{d}} d \eta_d \int_0^{2\pi} d\phi_{d} \: P(\hat{\D\vx} | \veta_d)\: P(\hat{\veta} | \veta, \veta_d)
\,.
\ea
In the second line we have used that the probability factorizes into
one for the centroid shift, which only depends on the orientation $\veta_d$,
and a second factor for the shear estimate.  We assume a Gaussian error
on $\hat{\D\vx}$, so that
\be
P(\hat{\D\vx} | \veta_d) = N_2\left[\hat{\D\vx} - \D\vx_{\rm true}(\veta_d), \:\s_{\D\vx}\right]\,.
\ee
Here, $N_2$ denotes the bivariate normal distribution and $\s_{\D\vx}$ is the
square root of the variance of each component of $\D\vx$ [given by the
observational error on the centroid shift, in our case \refeq{centr_noise}].  
Further, $\D\vx_{\rm true}(\eta_d)$ denotes the true (mean) centroid
shift of a galaxy oriented by $\theta,\phi$ related to $\veta_d$ through
\refeq{shearorient}.  On the other hand, we neglect observational errors
on the second moments of the galaxy image, so that
\be
P(\hat{\veta} | \veta, \veta_d) = \d_D^{(2)} [ \hat{\veta} \ominus (\veta \oplus \veta_d) ]\,.
\label{eq:Petahat}
\ee
That is, the shear estimated from the observed second moments is exactly
equal to the shear-space composition of the orientation $\veta_d$ and the assumed 
lensing shear $\veta$.  Of course, these do not in general have the same
azimuthal angle.  It is straightforward to generalize \refeq{Petahat} to
allow for a finite uncertainty in the shear estimate $\hat{\veta}$.  

Performing the integral over $\veta_d$ straightforwardly yields
\ba
P(\veta | \hat{\D\vx},\,\hat{\veta}) 
& = \frac1{\N'} P(\eta) \label{eq:post2}\\
 \times\:& \left\{ e^{-\eta_d} N_2\left[\hat{\D\vx} - \D\vx_{\rm true}(\veta_d), \:\s_{\D\vx}\right] \right\}_{\veta_d = \hat{\veta} \ominus \veta}
\,.
\nonumber
\ea
If there is no information on the centroid shift (equivalent to taking
$\s_{\D\vx} \to \infty$), and we assume a flat prior, then
\be
P(\veta | \hat{\D\vx},\,\hat{\veta})  \propto \cos\theta(\veta,\hat{\veta})\,,
\label{eq:postlowSN}
\ee
where $\theta(\veta,\hat{\veta})$ is the orientation angle of a galaxy that,
when lensed by
$\veta$, yields moments $\hat{\veta}$.  This is the expected PDF of shape
noise for perfect circular disks.

For our studies, $\hat{\veta}$ will be assumed to correspond to a disk
oriented by some fixed fiducial angles $\theta_0,\,\phi_0$ (that is, we assume no lensing), and we will
let $\phi_0 = 0$ without loss of generality.  We then define $\eta_0$ through
$\exp(-\eta_0) = \cos\theta_0$.  
Further, let $\varphi$ denote the azimuthal angle of $\veta$.  Then,
$-\veta$ has azimuthal angle $\varphi + \pi/2$ (a sign flip in the shear
corresponds to rotation by 90 degrees).  We then obtain
\ba
\veta_d =\:& \hat{\veta} \oplus (-\veta) \qquad \Leftrightarrow \qquad \vs
\cosh \eta_d =\:& \cosh \eta_0 \cosh \eta - \sinh \eta_0 \sinh \eta \cos (2\varphi) \vs
\sin (2\varphi_d) =\:& - \frac{\sinh \eta}{\sqrt{\cosh^2 \eta_d - 1}} \sin (2\varphi)\,.
\ea
Note that $\varphi=0$ implies $\varphi_d = 0$.  This is because the sum
of two aligned shears is in the same direction.  For
$\varphi \neq 0$, we pick the solution that connects continuously to 
$\varphi=0= \varphi_d$.

In order to solve for $\theta_d$, we need to pick a branch of the quadratic solution.  
This is easily done using $\cosh \eta_d \geq 1$ and $\cos\theta_d \in [0,1)$.  
We obtain
\be
\cos\theta_d = \cosh\eta_d - \sqrt{\cosh^2 \eta_d - 1}\,.
\ee
For a fixed value of $\theta_0$, it is then straightforward to map out the
posterior \refeq{post2} in the $\eta$ plane.  

\begin{figure}
\centering
\includegraphics[width=0.49\textwidth]{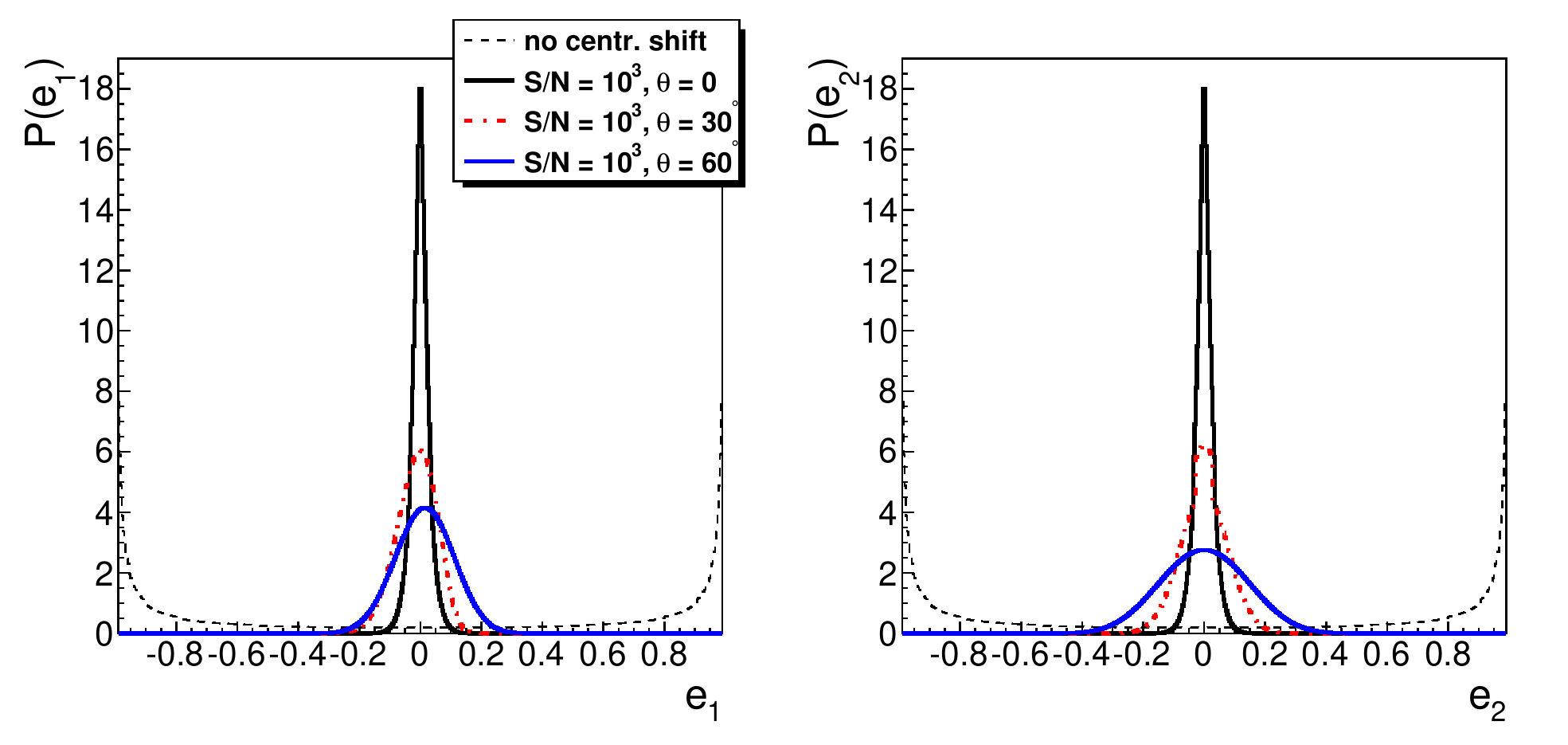}
\caption{Marginalized posterior for the ellipticity components $e_1,\,e_2$ for
$\nu = 1000$ and different inclination angles $\theta$.  In our coordinate
convention ($\varphi=0$), $e_2$ is the component unaffected by the inclination
angle.  The thin black line
shows the posterior without any orientation information, equivalent to
letting $\nu\to 0$ in our calculation.
\label{fig:postmarg}}
\end{figure}

However, in order to derive the shape noise predicted by this posterior, 
it is more practical to transform
from $\veta$ to the ellipticity (or distortion) $\v{e}$, defined through
\be
e_1 = \frac{I_{11} - I_{22}}{{\rm Tr}\:I_{ij}} ; \quad
e_2 = \frac{2 I_{12}}{{\rm Tr}\:I_{ij}} \,.
\label{eq:edef}
\ee
Note that $e_i \in [-1,1]$, and that at linear order, lensing changes the ellipticity
$\v{e}$ through
\be
\v{e} \to \v{e} + 2\bm{\gamma}
\label{eq:egamma}
\ee
where $\bm{\gamma}$ is the shear vector.  Following Eq.~(2.7) of \cite{bernstein/jarvis},
\be
\v{e}(\veta) = \tanh \eta \left(
\begin{array}{c}
\cos 2\varphi \\
\sin 2\varphi
\end{array}\right)\,,
\ee
where $\varphi$ is the azimuthal angle of $\veta$, so that 
\ba
P(\v{e} | \hat{\D\vx},\,\hat{\veta}) =\:& P(\veta(\v{e}) | \hat{\D\vx},\,\hat{\veta})
\left| \frac{\partial\veta}{\partial\v{e}}\right| \vs
=\:&  P(\veta(\v{e}) | \hat{\D\vx},\,\hat{\veta}) \left(1-e^2\right)^{-2}\,,
\ea
where
\be
\eta(e) = \frac12 \ln \left(\frac{1+e}{1-e}\right)\,.
\ee

\reffig{postmarg} shows the posterior for the two components $e_1,\,e_2$
of the residual 
ellipticity for different true orientation angles at fixed 
signal-to-noise $\nu = 1000$.  Clearly, the ellipticity is well
constrained, with sharper constraints for smaller inclinations.  This
is because degeneracies between orientation and lensing increase for 
finite inclinations.  The thin black line shows the posterior if no
centroid shift is measured.  The posterior is then very broad with a peak
at $e_i \to \pm 1$.  This is just the consequence of the uniform prior
on $\cos\theta$.  

\begin{figure}
\centering
\includegraphics[width=0.49\textwidth]{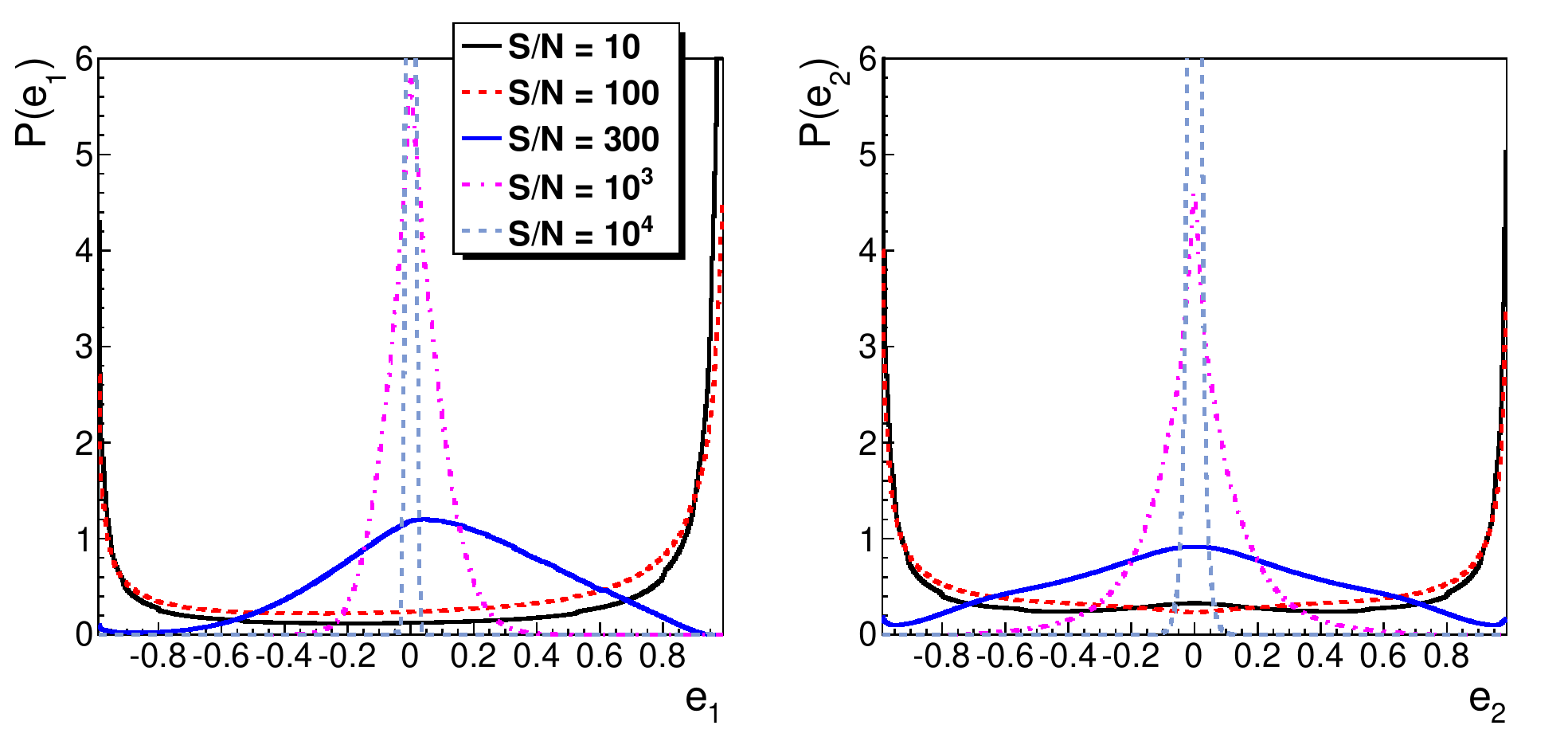}
\caption{Marginalized posterior for the ellipticity components $e_1,\,e_2$ 
averaged over inclination angles for different values of signal-to-noise.
\label{fig:post_e}}
\end{figure}

\reffig{post_e} shows the posterior for the ellipticity components 
averaged over inclination angle $\theta_0$.  The results are shown for different
values of the image signal-to-noise.  There is a rapid transition between values
$\nu \sim 100$, up to which the distribution of residual ellipticity is 
broad, and $\nu \sim 10^3$, when the orientation of the galaxy is
well constrained and the residual ellipticity follows a narrow distribution
centered around zero.  

\begin{figure}
\centering
\includegraphics[width=0.49\textwidth]{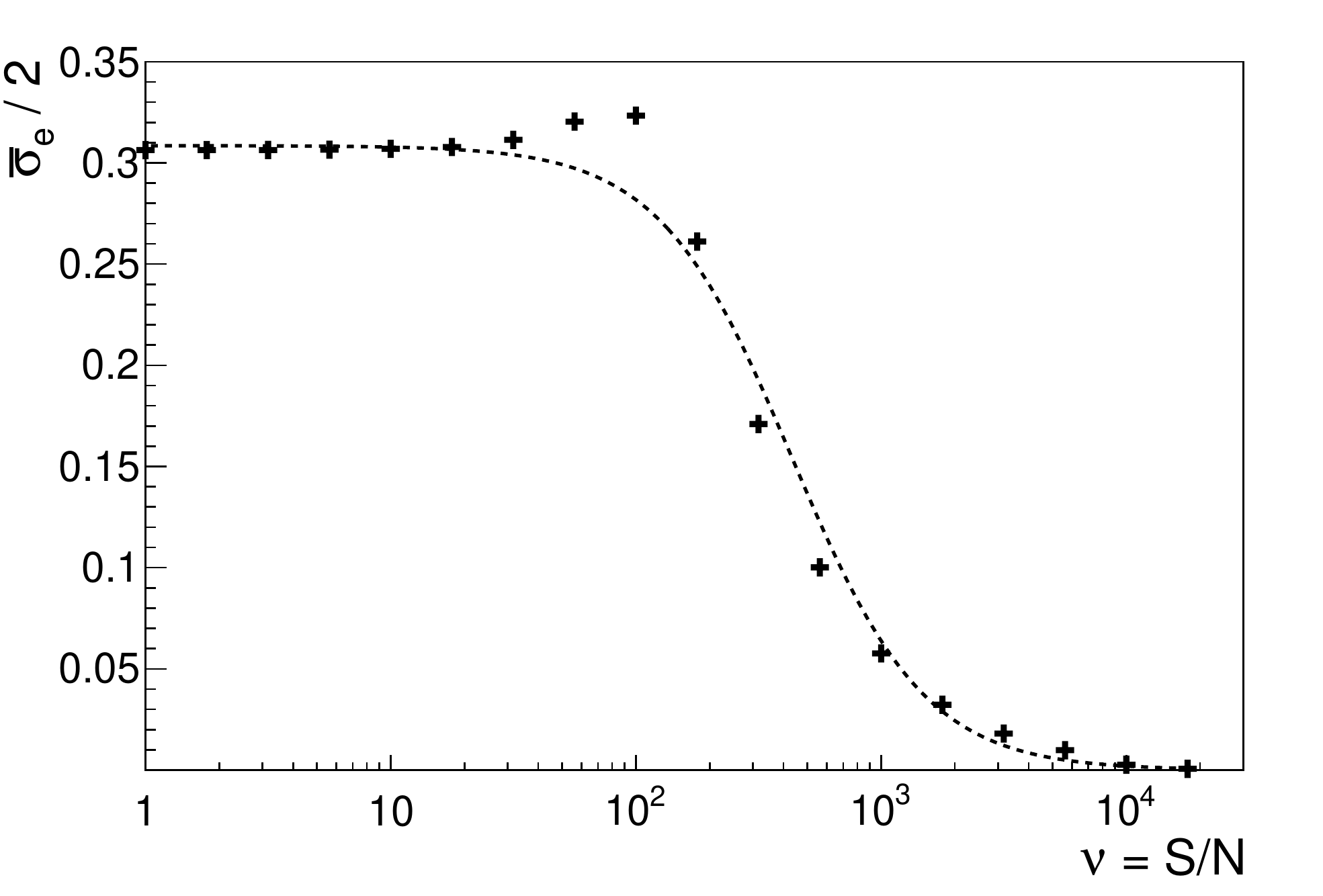}
\caption{RMS ellipticity averaged over $\theta_0$ following \refeq{sigmae_weighted} as function of signal-to-noise.  We have divided the RMS by 2 since that gives the 
shape noise contribution to the RMS of each shear component.   
The line shows the fit given in \refeq{sigres_fit}.}
\label{fig:RMSe_vs_SN}
\end{figure}

In the end, we would like to estimate the residual shape noise after the
centroid shift has been optimally taken into account.  This is
approximately quantified by the RMS of the posterior for the two
ellipticity components, specifically
\be
{\rm RMS}(e_i) = \left[\int_{-1}^1 d^2 \v{e}\, (e_i)^2 P(\v{e} | \hat{\D\vx},\,\hat{\veta})\right]^{1/2}\,,
\ee
where we have used that both components have vanishing expectation value
(unlike the magnitude $|\v{e}|$).  
In order to estimate the residual shape noise, we use the quantity $\sigma_e$
defined through
\be
\sigma_e = \frac1{\sqrt{2}} \left( \left[{\rm RMS}(e_1)\right]^2 + \left[{\rm RMS}(e_2)\right]^2 \right)^{1/2}\,,
\ee
which is the average noise per component of the ellipticity.  
As shown in \reffig{postmarg}, this quantity depends on the orientation angle of the
galaxy.  In order to estimate the effective shape noise for an ensemble
of galaxies with random orientations, we perform an inverse-variance weighting
to obtain
\be
\bar\sigma_e = \left[\int_0^1 d\cos\theta_0\: \sigma_e^{-2}(\theta_0)\right]^{-1/2}\,.
\label{eq:sigmae_weighted}
\ee
This is not exactly what one would do in reality, since in practice one has to
use the centroid shift and/or shape of the galaxy to estimate $\theta_0$.  
Further, the posterior for $e_i$ is highly non-Gaussian, especially for
small signal-to-noise, so that \refeq{sigmae_weighted} is not optimal.  However, we take 
it as a first approximation.  
The result is shown as function of galaxy signal-to-noise in \reffig{RMSe_vs_SN}, where we have divided $\bar\sigma_e$ by two in order to yield the effective shape noise $\sigma_\gamma$.    
It shows the expected behavior: for low signal-to-noise values where the centroid shift
is not detectable observationally, the RMS of the residual ellipticity 
approaches a constant corresponding to the usual shape noise for randomly
oriented disk galaxies.  Once
the centroid shift can be measured, the residual RMS drops rapidly.  
In the intermediate regime ($\nu \sim 100$), we see that the RMS
actually increases for increasing signal-to-noise.  This is due to the deficiencies
of the simple weighting scheme in \refeq{sigmae_weighted}; indeed, when
performing a uniform weighting over orientation, this feature goes away.  
Note that for an ellipticity RMS better than $0.1$, achieved
at signal-to-noise values of order a thousand, we expect
departures from a perfect thin disk to become relevant \citep{ryden:04}.

\subsection{Impact on shear surveys}

We have seen that the centroid shift can significantly reduce the
residual shape noise for disk galaxies at high signal-to-noise.  In the following,
we will use
\be
\sigma_{\rm int, res} = \frac12 \bar\sigma_e
\label{eq:siggamma}
\ee
as estimate for the residual shape noise, i.e. the RMS in each component
of the shear, where the factor of $1/2$ comes from \refeq{egamma}.  The
question then arises to what extent this reduces the noise overall in
shear surveys.  Note that this reduction only applies to disk
galaxies, which make up the majority of but not all galaxies in imaging surveys.  Thus, when we talk about the number density of galaxies in the following
we will always mean the number density of disk galaxies.
In order to estimate the reduction in noise, we define the Fisher information delivered by galaxies
within a logarithmic interval in the signal-to-noise $\nu$:
\be
\frac{dF}{d\ln\nu} = \frac{dn_g}{d\ln \nu} \sigma_\gamma^{-2}(\nu)\,,
\ee
where $dn_g/d\ln\nu$ is the number density of source galaxies per 
logarithmic $\nu$ interval, and $\sigma_\gamma$ is the effective shape noise
for galaxies of signal-to-noise $\nu$.  At fixed redshift, $dn_g/d\ln\nu$
roughly follows a Schechter function, i.e. it drops exponentially at
high $\nu$.  On the other hand, when averaged over a wide redshift
range, $dn_g/d\ln\nu$ approaches a power law.  Samples selected for
lensing measurements typically use a cut on photometric redshift
to exclude low-redshift galaxies which do not contribute to the
lensing measurement.  For this reason, we will assume an exponential
distribution,
\be
\frac{dn_g}{d\ln \nu} \propto \exp\left(-\frac{\nu}{\nu_*}\right)\,,
\label{eq:dngdlnnu}
\ee
up to a normalization constant which is irrelevant here.  
This choice is conservative, as it reduces the number of very high signal-to-noise
galaxies as compared to a power law distribution.  We have used the weak 
lensing shear sample of the COSMOS survey \citep{cosmosshear} to confirm that
$\nu_* \approx 60$ roughly describes the signal-to-noise distribution
for photometric redshifts $z \in [0.5, 1.5]$ (note that the values of $\nu$
in the catalog are only approximate, since they neglect pixel-level noise
correlations).  

\begin{figure}[t]
\centering
\includegraphics[width=0.49\textwidth]{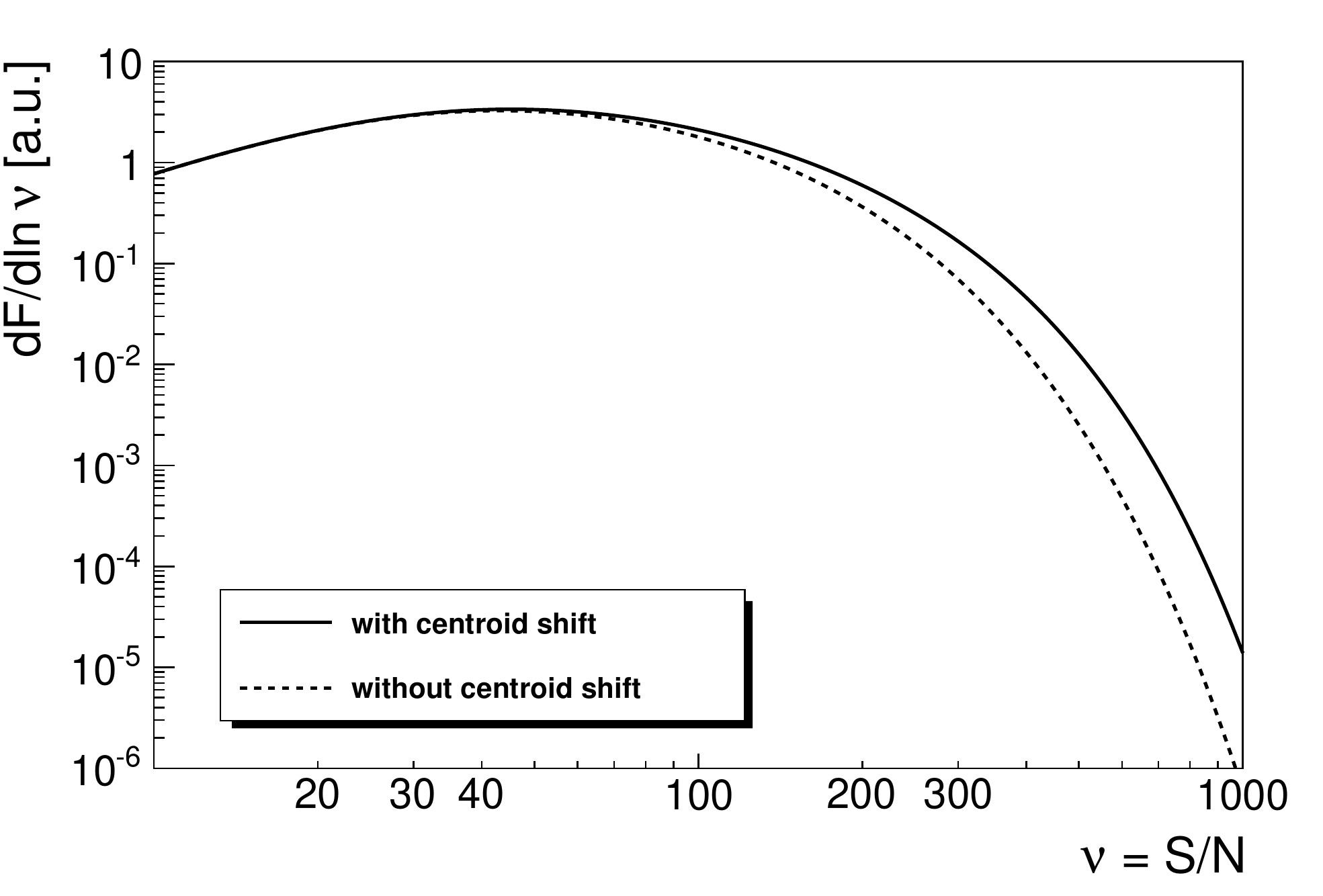}
\caption{
Information content of galaxies for shear estimation
per logarithmic interval in signal-to-noise $\nu$ without 
(dashed) and with (solid) using the centroid shift.}
\label{fig:ngal_vs_SN}
\end{figure}

For the shape noise, we assume 
\be
\sigma_\gamma^2(\nu) = \sigma_{\rm meas}^2(\nu) + \sigma_{\rm int, res}^2(\nu)\,,
\ee
where $\sigma_{\rm meas}$ is the contribution from measurement errors in the
galaxy shape, while $\sigma_{\rm int, res}$ is the residual shape noise due to
the imperfect measurement of the centroid shift and resulting uncertainty
in the orientation.  The latter is calculated from the results of the
previous section through \refeq{siggamma}.  
We assume that $\sigma_{\rm meas}$ is inversely proportional to $\nu$, specifically
\be
\sigma_{\rm meas}(\nu) = \frac{\nu_0}{\nu}\,,
\ee
and we adopt $\nu_0 = 10$ as fiducial value.  The scaling with $\nu^{-1}$
has been found to roughly describe the uncertainty of the image second moments
in the simulations of \cite{massey/etal:07} (\refapp{sigma_centroid}).  
The value of $\nu_0$ was simply chosen to yield a typical noise level 
expected in shear samples involving a signal-to-noise threshold of 20--40.
For convenience, we parametrize
$\sigma_{\rm int,res}$ through 
\be
\sigma_{\rm int,res}(\nu) = 0.31 \frac{(\nu/\nu_r)^{-p}}{1 + (\nu/\nu_r)^{-p}}\,,\quad \nu_r \approx 433\,, \quad p \approx 1.6\,,
\label{eq:sigres_fit}
\ee
which provides a reasonable fit (see \reffig{RMSe_vs_SN}).  
The results are shown in \reffig{ngal_vs_SN}.  The
centroid shift significantly increases the information content of high-signal-to-noise
galaxies.  For the fiducial values chosen here, it does not significantly
increase the total information on weak lensing shear, given by the 
integral under the curves in \reffig{ngal_vs_SN}, which is dominated by
galaxies with $20 \lesssim \nu \lesssim 100$.  This conclusion however depends
strongly on the amplitude of the centroid shift.  A centroid shift
greater by a factor of two, which could be possible in imaging surveys due to larger optical
depth of galaxies at higher redshift, already yields a significant enhancement in
the total information.  Finally, these results also depend on the
precise high-$\nu$ tail of the signal-to-noise distribution of galaxies, 
which will not be exactly exponential.

Note that for sufficiently high $\nu$, one should take into account the noise 
from the fact that galaxy disks are not perfectly circular.  However, this
will only become important for $\nu \gtrsim 10^3$.  One important caveat is
that we have assumed that the centroid
shift is only a function of inclination, and neglected scatter in the
dust content and scale height ratios from galaxy to galaxy.  Further
investigation, in particular using actual data, is warranted in order
to realistically assess the importance of the centroid shift for reducing the
shape noise in shear surveys.

\section{Probing the dust content of disk galaxies}
\label{sec:dust}

An alternative application of the centroid shift is to use the
measurement in conjunction with measured shapes to obtain the optical
depth due to dust extinction for a large sample of galaxies.  This could
be an interesting complement to existing studies of dust content which
are typically restricted to fairly small sample sizes.  

\begin{figure}[t!]
\centering
\includegraphics[width=0.49\textwidth]{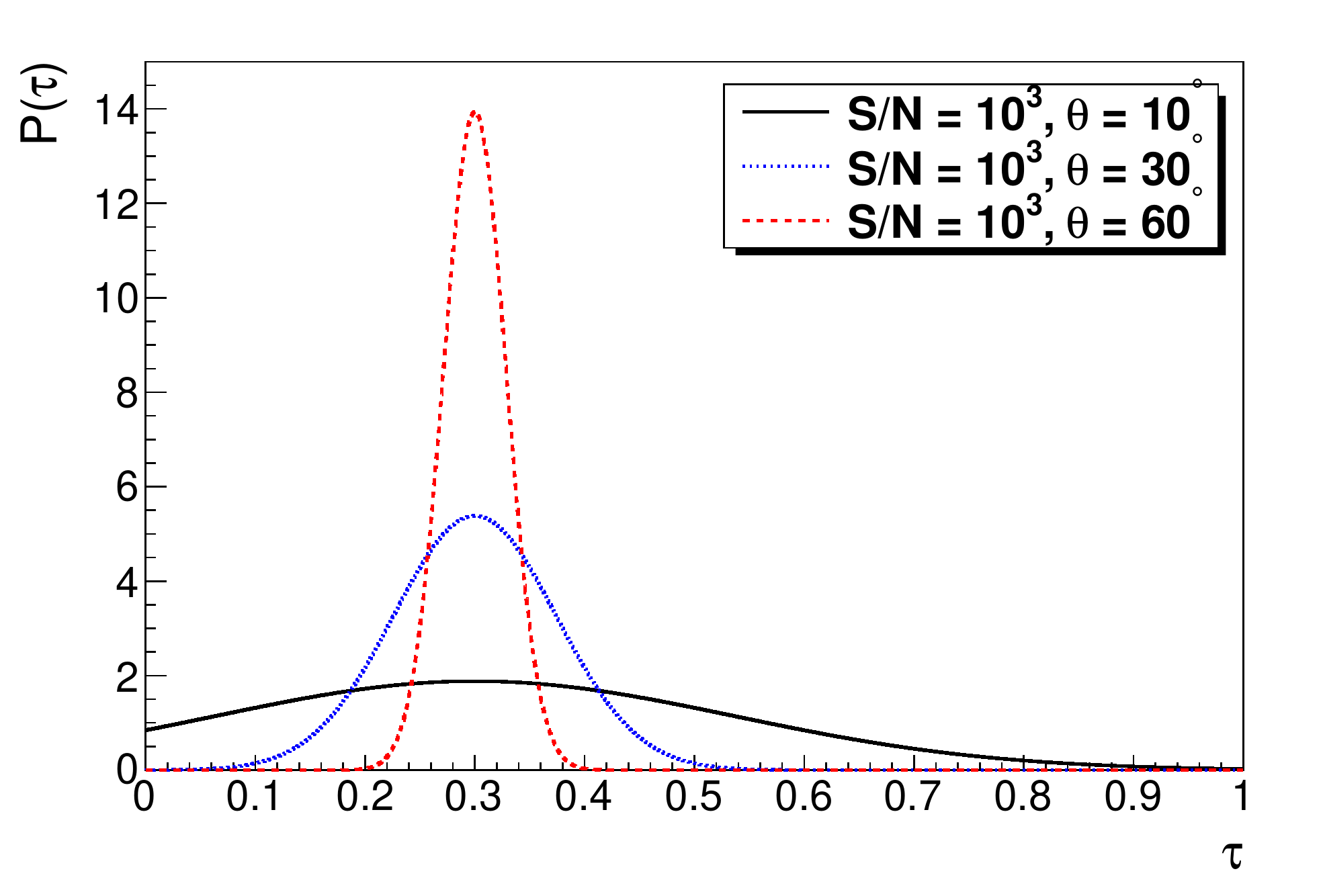}
\caption{Posterior for the relative face-on optical depth between bands
from which the centroid shift is measured, for
$\nu = 1000$ and different inclination angles $\theta$.  Our fiducial choice
is $\tau_{\rm true} = 0.3$.  Note that for $\theta\to 0$, there is no
constraint on the optical depth.
\label{fig:posttau}}
\end{figure}

The approach is the following: we now use the galaxy shape to constrain
the orientation of the galaxy, which in turn predicts the centroid shift 
assuming a value of the optical depth.  Comparing this to the measured
centroid shift allows us to constrain the optical depth.  
Let us write the actual centroid shift as
\be
\D\vx_{\rm true}(\tau,\theta) = \tau\:\D\vx_0(\theta)\,,
\ee
where $\tau = \tau_{{\rm fo},a} - \tau_{{\rm fo},b}$ is the difference in face-on optical depth 
between the two bands which are used to measure the centroid shift.  
$\D\vx_0(\theta)$ is assumed to be a universal function, although in reality this
will vary from galaxy to galaxy due to different dust distributions.  
Our goal is then to construct the posterior for $\tau$ given a measurement
of the centroid shift $\hat{\D\vx}$ and second moments on the sky $\hat I_{\rm sky}$.  As in the previous section, we will parametrize $\hat I_{\rm sky}$ through
the conformal shear vector $\hat{\veta}$.  Thus, our goal is to calculate
$P(\tau | \hat{\D\vx},\hat{\veta})$.  
In principle, one should marginalize over the weak lensing shear when
estimating the orientation from $\hat{\veta}$.  However, since the lensing
effect is small and we are only interested in a rough estimate, we will assume
that there is no lensing; in the framework of the derivation in the previous section, we adopt
a prior of $\d_D^{(2)}(\veta)$.  This yields
\ba
P(\tau | \hat{\D\vx},\hat{\veta}) =\:& \frac1{\N} P(\hat{\D\vx},\hat{\veta} | \tau) P(\tau) \vs
=\:& \frac{P(\tau)}{\N} \int d\cos\theta \int d\phi\: P(\hat{\D\vx}, \hat{\veta}|\tau,\theta,\phi)\,,
\nonumber
\ea
where $\N$ is a normalization constant, $P(\tau)$ is the prior on $\tau$, 
and $\theta,\phi$ are the true orientation angles of the galaxy.  
As before, we neglect the uncertainty on $\hat{\veta}$, assuming that
the second moments are very well measured, and adopt a Gaussian distribution
for the observed centroid shift.  This yields
\ba
P(\tau | \hat{\D\vx},\hat{\veta})
& = \frac1{\N'} P(\tau) \label{eq:posttau}\\
 \times\:& \left\{ e^{-\eta} N_2\left[\hat{\D\vx} - \D\vx_{\rm true}(\tau, \veta), \:\s_{\D\vx}\right] \right\}_{\veta = \hat{\veta}}
\,.
\nonumber
\ea
Let us assume that the expectation value of $\hat{\D\vx}$ is $\D\vx_{\rm true}(\tau_{\rm true}, \veta_{\rm true})$, where $\tau_{\rm true}$ is the actual optical depth.  
One can easily see that \refeq{posttau} becomes
\ba
P(\tau | \hat{\D\vx},\hat{\veta}) 
& = \frac1{\N''} P(\tau) \cos\theta\label{eq:posttau2}\\
 \times\:& \exp\left[- \left(\frac{\D\vx_0(\theta)}{\s_{\D\vx}}\right)^2 (\tau - \tau_{\rm true})^2\right]\,,
\ea
where $\theta$ is the orientation angle estimated from $\hat{\veta}$, which
under our assumptions is the true orientation angle.  Thus, at fixed
orientation angle the optical depth is Gaussian-distributed with RMS
\be
{\rm RMS}(\tau) = \frac{\s_{\D\vx}}{\D\vx_0(\theta)}\,.
\label{eq:RMStau}
\ee
In particular, for $\theta\to 0$ the variance becomes infinite, since the
centroid shift vanishes for any $\tau$.  This is illustrated in \reffig{posttau}, where we show the posterior for $\tau$ at fixed signal-to-noise for
different values of $\theta$.  

We can now perform an inverse-variance
weighting over orientation angles, in analogy with \refeq{sigmae_weighted}.  
The result can be simply cast into the form
\be
\bar\sigma_\tau = 0.29 \left(\frac{\nu}{100}\right)^{-1}\,.
\ee
For galaxies with signal-to-noise greater than a few hundred, $\tau$
can thus be measured very accurately.  In order to roughly estimate
the number density of galaxies in state-of-the-art imaging surveys
such as CFHTLenS \citep{cfhtlens}, DES \citep{DES:05}, or HSC 
\citep{Miyazaki/etal:12}
that will yield dust estimates through centroid shift measurements, we assume
a number density of $\sim 10~\rm arcmin^{-2}$ above $\nu \sim 20$, which
leads to an estimate of $\sim 0.1~\rm arcmin^{-2}$ for galaxies with
$\nu \gtrsim 200$ using our exponential ansatz in \refeq{dngdlnnu}.  
This corresponds to 360 galaxies per square degree with detectable
centroid shift.  Taking into account the considerable uncertainties in this measurement,
an expectation of several hundred galaxies per square degree seems reasonable,
which, for surveys covering thousands of square degrees, yields  
dust extinction estimates for several hundred thousand galaxies.

\section{Conclusions}
\label{sec:concl}

We have shown that dust extinction leads to a shift of the centroid of 
galaxy images in different bands, focusing in particular on disk galaxies.  
This shift is a well-defined function of inclination of the disk, with
amplitude mainly determined by the intrinsic, face-on optical depth
in each band, and the relative scale heights and lengths of opacity
and emissivity in the galaxy.  We estimated that this centroid shift
should be detectable for galaxies with signal-to-noise values greater
than of order 100.  In state-of-the-art imaging surveys, this translates
to number densities of \ngal galaxies per square degree with measurable
centroid shift.  This centroid shift could be interesting
for two purposes.

First, given independent (statistical) constraints on the dust optical
depth, the centroid shift can be used to infer galaxy orientations.  
The unknown orientation is the main reason for non-circular apparent
shapes of disk galaxies, and thus the main source of shape noise
in weak lensing shear surveys.  For high signal-to-noise galaxies,
the centroid shift can thus reduce shape noise and significantly
increase their importance for estimating shear.  We have found that
for our fiducial assumptions about typical optical depths, the increase
in overall information on shear is modest.  However, should dust extinction
increase towards higher redshift (where most source galaxies in shear
surveys reside), then this approach could yield a significant increase
in lensing information extractable from imaging surveys.  A caveat
to this is that we have not included scatter in optical depth and scale
height ratios which determine the normalization of the centroid shift-inclination
relation for a given galaxy.  We believe however that this application of
the centroid shift warrants further study.

A complementary application of this effect is to neglect lensing and use 
the observed galaxy shapes as a measure of their orientation.  Then,
the centroid shift is a new, independent probe of dust extinction which
is directly applicable to a large number of galaxies (we have estimated
hundreds of thousands of galaxies in state-of-the-art imaging surveys).  
Together with
modeling of galaxy SEDs, this should provide crucial constraints breaking
some of the degeneracies
present in dust modeling, e.g. the distribution of dust and orientation.

Finally, and perhaps most importantly, one of the first steps should be to look 
for this effect in existing data.  Our estimates show that it should easily be detectable
in currently available samples such as from SDSS or CFTHLenS.\\

\emph{Acknowledgments:} We would like to thank
Jim Bosch,
Bruce Draine,
Simone Ferraro,
Brandon Hensley,
Eric Huff,
Guinevere Kauffmann,
Alexie Leauthaud,
Robert Lupton,
Rachel Mandelbaum, 
Richard Massey,
Hironao Miyatake,
David Spergel and
Michael Strauss
for helpful discussions and comments.

\begin{widetext}
\appendix

\section{Analytical estimate of centroid offset}
\label{app:smallangle}

In this appendix we derive an analytical solution to \refeq{Isolobs} in the
limit that $z_{\rm gf}\tan\theta/r_{\rm gf}$ is a small parameter.  This limit assumes that
the radial distance traveled by the light ray while within the region where
$\eps$ and $\k$ are non-negligible is much smaller than the radial distance
at which it originated.  Clearly, this condition can be satisfied by 
making $\theta$ sufficiently small, but it can also be satisfied at finite
$\theta$ when the disk scale height of the galaxy is much smaller than its
scale length, since the integrand in \refeq{Isolobs} is then only relevant for
$|z_{\rm gf}| \ll r_{\rm gf}$.  Conversely, for sufficiently large inclinations $\theta$ this
expansion will break down.  

Throughout we will work in galaxy-frame cylindrical coordinates $r_{\rm gf}, \varphi_{\rm gf}, z_{\rm gf}$, and for clarity
we will drop the subscript ``gf'' in the remainder of this appendix.  
We expand \refeq{Isolobs} to linear order in $z\tan\theta/r$, yielding
\ba
I_{\rm obs}(r, \varphi; \theta) = \frac1{\cos\theta}\left[ I_0(r; \theta) + \cos\varphi\, I_1(r; \theta)\right]
\ea
where we have defined
\ba
I_0(r; \theta) =\:& \int_{-\infty}^\infty dz \: \eps(r, z) \exp\left[-\frac1{\cos\theta}\int_{z}^{\infty} \k(r, z') dz' \right]
\label{eq:I0}\\
I_1(r; \theta) =\:& 
\tan\theta
\int_{-\infty}^\infty dz \left[ z \frac{\partial \eps(r,z)}{\partial r} 
- \frac{\eps(r,z)}{\cos\theta} \int_z^{\infty} dz'\:z' \frac{\partial\k(r,z')}{\partial r}
\right]
\exp\left[-\frac{1}{\cos\theta}\int_{z}^{\infty} \k(r, z') dz' \right] \,.
\label{eq:I1}
\ea
$I_0$ is the leading order intensity distribution of the galaxy in this 
expansion, which is independent of $\varphi$ and hence does not show a centroid
shift.  
The centroid shift is induced by the second term, \refeq{I1}, which is
proportional to $\cos\varphi$.  We define the centroid of the observed
intensity $\vx_c$ through, 
\be
\vx_c = \left[\int r dr \:w(r)\int d\varphi\: I_{\rm obs}\right]^{-1}
\int_0^{\infty} r dr \:w(r) \int_0^{2\pi} d\varphi\: 
\left(\begin{array}{cc}
r \cos\varphi\\
r \sin\varphi
\end{array}\right) I_{\rm obs}(r,\varphi; \theta)\,,
\ee
where we have allowed for a weighting function $w(r)$ (see \refapp{centroid_estimate}).  
\refeqs{I0}{I1} then immediately yield at leading order
\ba
x_c^1 =\:& \frac12 \left[\int r dr \:w(r) I_0(r;\theta)\right]^{-1}
\int_0^{\infty} r^2 dr \:w(r) I_1(r;\theta) ; \quad
x_c^2 = 0\,.
\label{eq:xc1}
\ea
As expected, the centroid shift is aligned with the $x$ axis for the 
coordinate system adopted.  
We thus need to estimate the relative magnitudes of $I_1(r;\theta)$ and
$I_0(r;\theta)$.  

We can make the expressions more transparent by defining the face-on integrated opacity or ``partial optical depth'' $\tau_{\rm fo}(r,z) = \int_z^\infty dz'\:\k(r,z')$ and the
scale lengths
\be
r_\eps^{-1}(r) \equiv -\frac{\partial\ln \eps(r,z\simeq0)}{\partial r}; \quad
r_\k^{-1}(r) \equiv -\frac{\partial\ln \k(r,z\simeq0)}{\partial r} \,,
\ee
which are positive if emissivity and opacity increase towards
the center of the galaxy.  Further, we define an effective scale
height for the emissivity $\hat{z}_\eps$ through 
\ba
\hat{z}_\eps(r; \theta) \equiv\:& \frac{1}{I_0(r; \theta)}\int_{-\infty}^\infty dz\:z\:\eps(r,z) e^{-\tau_{\rm fo}(r,z)/\cos\theta} \,.
\ea
Note that (for a galaxy symmetric under $z\to -z$) $\hat{z}_\eps$ vanishes unless
$\tau_{\rm fo}\neq 0$.  
Similarly, we define an effective opacity scale height,
\ba
\hat{z}_\k(r; \theta) \equiv\:& \frac{1}{I_0(r; \theta) \cos\theta}\int_{-\infty}^\infty dz\:\eps(r,z) e^{-\tau_{\rm fo}(r,z)/\cos\theta} 
\int_z^\infty dz'\:z'\k(r,z')\,.
\ea
In the following, we will omit the $r$ and $\theta$ dependence of 
$r_\eps,\,r_\k,\,\hat{z}_\eps,\,\hat{z}_\k$ for clarity.  At leading order,
we then have 
\be
\frac{I_1(r; \theta)}{I_0(r; \theta)} = \tan\theta \left(\frac{\hat{z}_\k}{r_\k}
- \frac{\hat{z}_\eps}{r_\eps} \right) \,.
\label{eq:DIgen}
\ee
From \refeq{xc1} we see that, modulo a radial weighting, this ratio determines
the fractional centroid shift.

As an example, consider the case where the emissivity is located at $z=0$,
while the absorbing material (dust) follows a more extended distribution,
with $r_\eps \sim r_\kappa$.  Then, $\hat{z}_\eps \ll \hat{z}_\k$.  Thus, \refeq{DIgen} becomes
\be
\frac{I_1(r; \theta)}{I_0(r; \theta)} = \tan\theta \frac{\hat{z}_\k}{r_\k}\,.
\ee
This says that the half of the galaxy closer to the observer ($x>0$) 
suffers less absorption than the other half further away.  
In the opposite case, where a distribution of emissivity surrounds a dust
screen at $z=0$, we have
\be
\frac{I_1(r; \theta)}{I_0(r; \theta)} = -\tan\theta \frac{\hat{z}_\eps}{r_\eps}\,.
\ee
In this case, more emissivity contributes on the side of the galaxy
that is further away from the observer, leading to a centroid shift in the 
negative $x$ direction.  
In general some cancellation can occur between the two competing effects,
although note that the effect will be non-zero for an identical distribution
of $\eps$ and $\k$ due to the different weighting involved in $\hat{z}_\eps$ and
$\hat{z}_\k$ (unless of course both $\hat{z}_\eps$ and $\hat{z}_\k$ vanish).  We thus already
see the two main factors which determine the centroid shift: the optical
depth $\tau \propto \int dz\,\k$, and the ratio $z/r$ of typical scale 
height to scale length of the galaxy.

We can now specialize to the case of an exponential disk, \refeq{exponentialdisk}.  
For simplicity, we will also linearize the result in $\tau_{\rm fo}$, which 
given $\tau_{\rm fo} \sim 0.2-0.5$ is sufficiently accurate for this approximate
estimate.  Defining the ratio $c_z = z_\k/z_\eps$ of opacity to emissivity
scale heights, we obtain
\ba
\hat{z}_\eps =\:&\frac{\tau_{\rm fo}(r)}{\cos\theta} 
\frac{1 + 2 c_z}{(1+c_z)^3}\:z_\eps ; \quad
\hat{z}_\k = \frac{\tau_{\rm fo}(r)}{\cos\theta} \frac{c_z^2 (2 + c_z)}{(1+c_z)^2}\:z_\eps\,,
\label{eq:hatzexpo}
\ea
which using \refeq{DIgen} yields
\ba
\frac{I_1(r; \theta)}{I_0(r; \theta)} =\:& \tan\theta\: \tau_{\rm fo}(r)
\left[\frac{r_\eps}{r_\k} \frac{c_z^2 (2 + c_z)}{(1+c_z)^2}
-  \frac{1 + 2 c_z}{(1+c_z)^3}\right]
\frac{z_\eps}{r_\eps}
\,.
\ea
Here we have dropped a $1/\cos\theta$ factor which converts the face-on
opacity to the actual observed opacity, since this is 1 at leading order 
in our expansion in $\tan\theta$.  

Finally, performing the radial weighting in \refeq{xc1}, we obtain
\be
\frac{|\vx_c|}{r_w} = \frac{|x_c^1|}{r_w} = \frac12 \tan\theta \left(\frac{\<\hat{z}_\k\>_{r w}}{r_\k}
- \frac{\<\hat{z}_\eps\>_{rw}}{r_\eps} \right)\,,
\ee
where the weighted effective radius of the galaxy is
\be
r_w = \left[\int r dr \:w(r) I_0(r;\theta)\right]^{-1}
\int_0^{\infty} r^2 dr \:w(r) I_0(r;\theta)\,,
\label{eq:rw}
\ee
and the radial weighting of the scale heights is defined through
\be
\< f(r) \>_{r w} \equiv \left[\int r^2 dr \:w(r) I_0(r;\theta)\right]^{-1}
\int r^2 dr \:w(r) f(r)\: I_0(r;\theta)\,.
\ee
Using the results for the exponential disk \refeq{hatzexpo}, this simply
becomes
\be
\frac{|\vx_c|}{r_w} =
\frac12 \tan\theta \<\tau_{\rm fo} \>_{rw}
\left[\frac{r_\eps}{r_\k} \frac{c_z^2 (2 + c_z)}{(1+c_z)^2}
-  \frac{1 + 2 c_z}{(1+c_z)^3}\right]
\frac{z_\eps}{r_\eps}\,,
\label{eq:xc2}
\ee
which involves the radially weighted face-on optical depth.  
Inserting the parameters from \refeq{diskpar}, this becomes
\be
\frac{|\vx_c|}{r_w} \simeq 0.04
\tan\theta \<\tau_{\rm fo} \>_{rw}\,.
\label{eq:xc3}
\ee
This result explains both the behavior with inclination angle of the
centroid shift determined using the radiative transfer calculation as well
as the fact that the centroid shift is largest in the innermost regions:  
choosing a weighting function $w(r) = \Theta(r_{\rm max}-r)$ which only includes 
the regions $r < r_{\rm max}$, we see that $\<\tau_{\rm fo}\>_{rw}$ will decrease
with increasing $r_{\rm max}$, matching the trend seen in \reffig{shift_threshold}.

For this paper, we adopt a weighting function corresponding to
$r_{\rm max} = r_{\rm eff}/2$.  In terms of the literature value $\tau_{\rm fo}$,
we then have $\<\tau_{\rm fo}\>_{rw} \approx 1.6 \tau_{\rm fo}$ from the
radiative transfer results, while $r_w \approx r_{\rm eff}$ , which leads to
\be
\frac{|\vx_c|}{r_{\rm eff}} \simeq 0.06
\tan\theta\: \tau_{\rm fo}\,.
\label{eq:xcf}
\ee

As a final note, we point out that dust extinction also modifies the
galaxy's observed second moments, i.e. its shape.  However, this 
effect is suppressed by another power of the small parameter $z \tan\theta/r$, 
and hence less likely to be detectable than the centroid shift.

\section{Optimal estimation of centroid offset}
\label{app:centroid_estimate}

\reffig{shift_threshold} shows that the centroid shift due to dust extinction is
not uniform over the galaxy image, but rather is most prominent in the
inner regions.  For this reason, we derive an approximation to the optimal
estimation of the image centroid for the purpose of measuring the
extinction-induced centroid offset.  Throughout we neglect point-spread-function effects,
i.e. we assume that the galaxy is perfectly resolved.

Let us divide the galaxy image into approximately elliptical contours of 
constant surface brightness (isophotes).  
A general weighted estimate of the centroid
of the galaxy image is then given by
\be
\vx_w = \frac{\sum_{\rm pixels} w(\vx_i) \vx_i}{\sum_{\rm pixels} w(\vx_i)}
= \frac{\int r dr\:w(r) \mu(r) \int_0^\pi d\phi \:\vx(r,\phi)}{\pi\int r dr\:w(r)\mu(r)}\,,
\ee
where $\mu(r)$ is the surface brightness, and in the second equality we 
have gone to the continuum limit and elliptical coordinates parametrized 
through $(r,\phi)$.  Our goal is to derive the weight function $w(\mu)$ that
optimally captures the centroid shift.

At fixed inclination $\theta$, we can write for the centroid shift of a 
single isophote $r$:
\be
\D\vx(r) = f(\mu(r)) \,\tau_{\rm fo}\,.
\ee
The optimal estimate for the centroid shift is then given by the optimal
estimate for $\tau_{\rm fo}$.  Let $\vx_c(r)$ denote the centroid of the isophote $r$,
and $\sigma_{\vx_c}(r)$ denote its error.  Assuming that the errors from
different isophotes are independent, the optimal estimate for $\tau_{\rm fo}$ is the 
inverse-variance weighted average of isophotes:
\be
\hat\tau_{\rm fo} = N \int dr \frac{f^2(\mu)}{\sigma_{\vx_c}^2(r)} \frac{\vx_c(r)}{f(\mu)}\,,
\ee
where $N$ is a normalization constant.
Thus, the desired weighting function is given by
\be
w(r) = \frac{f(\mu)}{\sigma_{\vx_c}^2(r)}\,.
\ee
We now need to estimate the error $\sigma_{\vx_c}$, specifically its scaling
with $r$.  Following the results of \refapp{sigma_centroid}, we assume that
\be
\frac{\sigma_{\vx_c}(r)}{{\rm max}(r, \rPSF)} \propto \nu^{-1} \propto \mu^{-1}(r)\,,
\ee
where we have assumed that the signal-to-noise $\nu$ of the elliptical annulus
is proportional to its surface brightness.  Clearly, this is only a 
rough approximation, but sufficient for our desired purpose.  We then obtain
\be
w(r) = \frac{f(\mu(r)) \mu^2(r)}{{\rm max}(r^2, \rPSF^2)} \,.
\ee
\reffig{shift_threshold} shows that $f(\mu)$ can roughly be approximated
as $\propto \mu^2$.  In this case, $w(r) \propto \mu^4(r) / {\rm max}(r^2, \rPSF^2)$, which for an exponential profile is very strongly peaked at small
radii.  Thus, for galaxies at cosmological redshifts the centroid estimate
will be dominated by the innermost resolved scales, i.e. $r \lesssim \rPSF$.

\section{Estimate of observational centroid shift uncertainty}
\label{app:sigma_centroid}

In order to estimate what accuracy the centroid measurement in an imaging 
survey has, we use image simulations prepared for the Shear Testing Programme
(STEP; \cite{massey/etal:07}).  These image simulations were
designed to match space-based imaging.  However, we expect that
the rough scaling derived from these simulations will also describe
ground-based imaging as long as the galaxies are well resolved.

We have several realizations
of noise for the same galaxy images, and we calculate the quantities
\be
\D_x = \frac{x_{c,1}-x_{c,2}}{d}; \quad
\D_y = \frac{y_{c,1}-y_{c,2}}{d}\,,
\label{eq:Dxy}
\ee
where $x_{c,i}$ correspond to the $x$-component of the centroid measured
in noise realization $i$, and analogously for $y$.  Here, $d$ is 
the weighted size following \cite{RRG}.  This is typically 
$\sim 0.6$ times the half-light radius.  We then calculate the RMS values of $\D_x$
and $\D_y$ and average them in quadrature (the mean has been found to be
consistent with zero, as expected).  This exercise can be done
for various subsamples of the set of simulated galaxy images.  Note that
the scatter in $\D_x,\,\D_y$ necessarily includes some contribution
from the scatter in galaxy sizes between different noise realizations.  
We use the size averaged over noise realizations to reduce this effect,
but note that our estimates of the centroid uncertainty will be
conservative in this respect.

\begin{figure}
\centering
\includegraphics[width=0.5\textwidth]{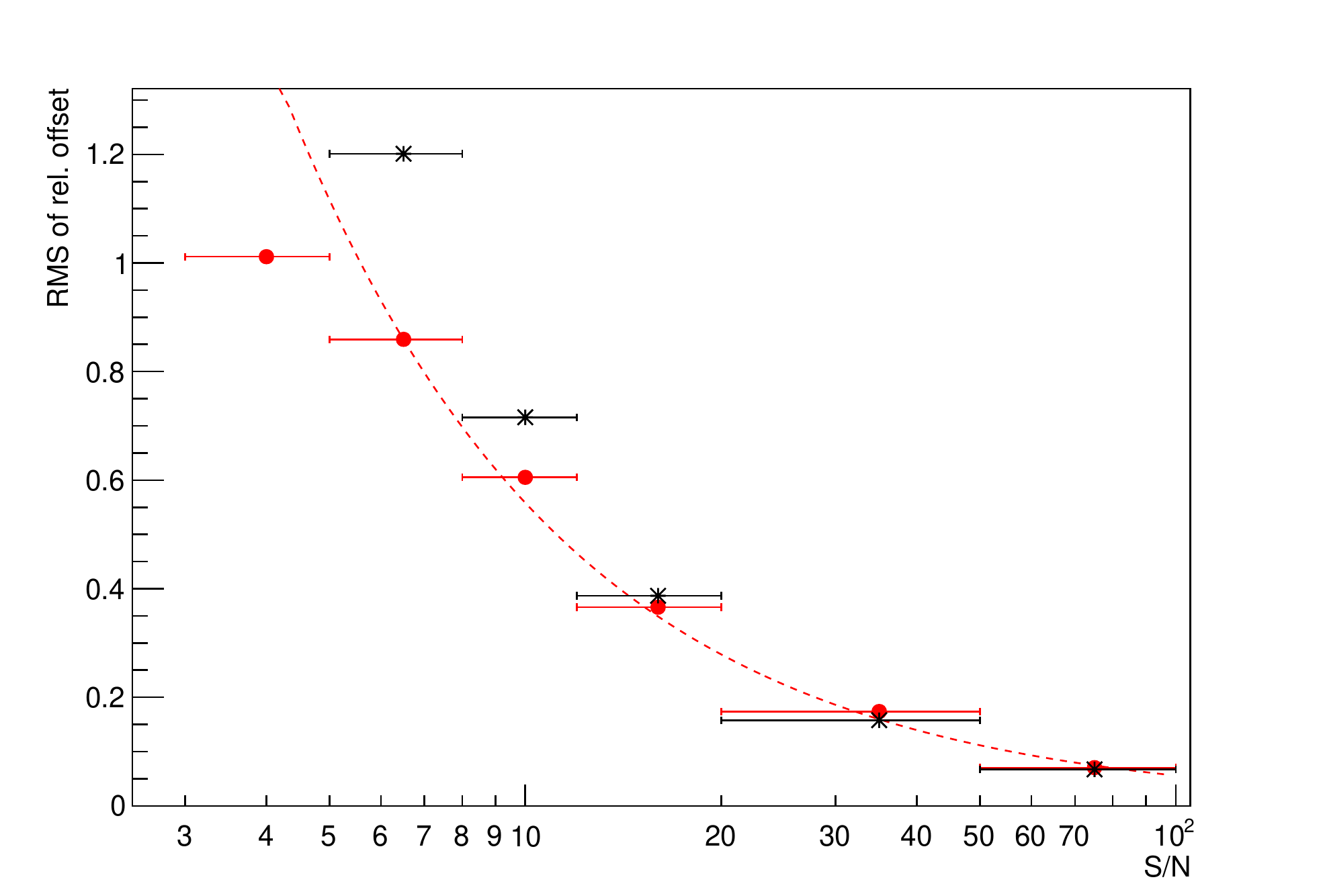}
\caption{RMS scatter in the relative centroid defined in \refeq{Dxy}
as a function of galaxy signal-to-noise, for two different samples of galaxies: 
those smaller than the median size (red, circles), and larger (black, stars).  
The dashed line indicates a simple $1/x$ fit.}
\label{fig:centroidRMS}
\end{figure}

We have found that the relative offset defined through \refeq{Dxy} is
quite insensitive to galaxy properties and mainly depends on the
signal-to-noise $\nu$ (while the absolute offset strongly depends on the size
of the image in addition to the signal-to-noise).  This is illustrated in 
\reffig{centroidRMS}, which
shows the RMS of $\D_x,\D_y$ as function of galaxy signal-to-noise, 
for two subsamples of different size.  There is some indication
that larger galaxies have larger scatter in the relative centroid
at fixed signal-to-noise, although the effect is not very large.  The dashed
line indicates a simple fit to the result for the smaller size sample:
\be
{\rm RMS}(\D_{x,y}) = \frac{5.6}{\nu}\,,
\ee
which translates to
\be
{\rm RMS}\left(\frac{x_c,\;y_c}{r_{\rm eff}}\right) = \frac{3.3}{\nu}\,,
\label{eq:centr_noiseA}
\ee
where $r_{\rm eff}$ is effective radius or the half-light radius.  
In other words, roughly speaking, the uncertainty in each of the centroid
components for a galaxy measured at $\nu=10$ is roughly one third of the
half-light radius.  A more accurate estimate would also consider only the
innermost resolved region of galaxies, but given the rough nature of
our forecasts we have not considered this here.  We have found
that the scatter in second moments of galaxy images obeys a similar rough
scaling with $\nu^{-1}$.

\end{widetext} 
\bibliographystyle{apsrev}
\bibliography{lensing}

\end{document}